\documentclass{emulateapj} 
\usepackage{enumerate,multirow}
\usepackage{amsmath,amssymb,graphicx,natbib,subfigure,dcolumn,bm,comment} 
\usepackage{amsmath}
\usepackage{amssymb} 
\usepackage{color}

\newcommand{\kms}{km s$^{-1}$}

\newcommand{\be}{\begin{equation}}
\newcommand{\ee}{\end{equation}}
\newcommand{\bi}{\begin{list}{\labelitemi}{\leftmargin=1em}\setlength{\itemsep}{-3pt}}
\newcommand{\ei}{\end{list}}

\newcommand{\beqn}{\begin{eqnarray}}
\newcommand{\eeqn}{\end{eqnarray}}

\shorttitle{Black holes in nuclear star clusters}
\shortauthors{Antonini and Rasio}

\begin{document}
\def\gap{\;\rlap{\lower 2.5pt
\hbox{$\sim$}}\raise 1.5pt\hbox{$>$}\;}
\def\lap{\;\rlap{\lower 2.5pt
 \hbox{$\sim$}}\raise 1.5pt\hbox{$<$}\;}

\def\eb#1{{\textcolor{blue}{[[\bf EB: #1]]}}} 
\def\mb#1{{\textcolor{red}{[[\bf MB: #1]]}}}

\newcommand\MBH{\rm MBH} 
\newcommand\NSC{NSC}
\newcommand\CMO{CMO} 

\title{Merging black hole binaries in galactic nuclei:
implications for advanced-LIGO detections} 

\author{Fabio Antonini and Frederic A. Rasio}
\affil{Center for Interdisciplinary Exploration and Research in Astrophysics (CIERA)
and Department of Physics and Astrophysics, 
Northwestern University,  Evanston, IL 60208}
 
\begin{abstract}
Motivated by the recent detection of gravitational waves from the black hole binary merger GW150914, we study the dynamical evolution of  (stellar mass) black holes  in galactic nuclei where massive star clusters reside.  
With masses of $\sim~10^7~M_\odot$
 and sizes of only a few parsecs,  nuclear star clusters are the densest stellar systems observed in the local universe and represent  a robust environment  where  black hole binaries can dynamically form, harden and merge.  We show that due to their large escape speeds, nuclear star clusters can retain a large fraction of their  merger remnants.
Successive mergers can then lead to significant growth and produce black hole mergers of several tens of solar masses similar to GW150914 and up to a few hundreds of solar masses, without the need  of invoking extremely low metallicity environments. We use a semi-analytical approach to describe the dynamics of  black holes in massive star clusters. 
Our models give a black hole binary merger rate  of  $\approx~1.5\rm~Gpc^{-3}yr^{-1}$ from nuclear star clusters, implying up to a few tens of possible detections per year with Advanced LIGO. Moreover, we find a local  merger rate of $\sim~1\rm~Gpc^{-3}yr^{-1}$  for high mass  black hole binaries similar to GW150914; a  merger rate comparable to that of similar binaries assembled dynamically in globular clusters. Finally, we  show that if all black holes receive  high natal kicks, $\gtrsim~50\rm~km~s^{-1}$, then nuclear star clusters will dominate the local merger rate of binary black holes compared to  either globular clusters or isolated binary evolution.
\end{abstract}

\section{Introduction}\label{intro}
On September 14, 2015  the Advanced  LIGO interferometer (aLIGO)
detected the event GW150914, which
has been interpreted
as the first direct observation of gravitational waves (GWs) from 
the inspiral and merger of a pair
of black holes \citep[BHs;][]{2016PhRvL.116f1102A}. The event
GW150914 was produced by two BHs
 with masses of $36^{+5}_{-4}\  M_\odot$
and $29^{+4}_{-4} \  M_\odot$ 
 (in the source frame), 
 at a redshift 
$z \approx 0.1$ assuming standard cosmology \citep{2016arXiv160203840T}.  
The detection of the gravitational-wave signal of GW150914  
has provided the first direct evidence that black holes with 
mass $\gtrsim 30\  M_\odot$ exist and that they can reside in binary systems.
Assuming that the source-frame binary BH merger rate is constant within the volume 
in which GW150914 could have been detected,
and that GW150914 is representative of the underlying binary BH  population, 
the BH-BH merger rate  is inferred to be $2-53 \rm\ Gpc^{-3}\  yr^{-1}$ in the comoving 
frame \citep{2016arXiv160203842A}.

\citet{2016ApJ...818L..22A} reviews
various channels  for the formation of BH binaries 
that can coalesce within a Hubble time thus becoming  potentially detectable
by aLIGO. These include dynamical formation in dense stellar 
environments \citep[e.g.,][]{2000ApJ...528L..17P,2009ApJ...692..917M,2010MNRAS.402..371B,2011MNRAS.416..133D,2015PhRvL.115e1101R}, and 
isolated binary evolution 
\citep[e.g.,][]{2002ApJ...572..407B,2012ApJ...759...52D,2015MNRAS.451.4086S,2016ApJ...819..108B,2016arXiv160302291D}. 
While most of the former literature focused on BH binaries 
forming in globular clusters (GCs), little attention has been devoted to the 
formation of such binaries in nuclear star clusters (NSCs).
Yet NSCs  have total stellar masses that are comparable to the whole
 stellar mass of the GC system for the galaxy, at least in the
 Milky Way, and
are the densest and most massive star clusters observed in the local
universe \citep[e.g.,][]{2004AJ....127..105B,2006ApJS..165...57C},
representing therefore 
a natural environment where dynamical processes can efficiently
 lead to the formation of BH binaries.
In this paper we consider the dynamical formation scenario, and 
 explore the contribution to the  BH binary merger rate from 
NSCs.

In a stellar cluster, stellar mass BHs  formed from the death of
massive stars, quickly
 segregate to the center through dynamical friction \citep{1943ApJ....97..255C,Spitzer}. In these 
high-density environments, BHs can efficiently interact with each other 
and dynamically form new binaries.
Such binaries  will subsequently harden through three-body interactions \citep{1975MNRAS.173..729H}.
Via such dynamical processes, 
GCs can produce a significant population of BH binaries
that, after being ejected from the cluster, will be able to merge in the local universe \citep[e.g.,][]{2016arXiv160202444R}.
Over the last years our understanding of the evolution of BHs in star clusters
has improved considerably thanks to numerical efforts \citep[e.g.,][]{2012MNRAS.422..841A,2013ApJ...763L..15M,2015MNRAS.450.4070W,2015ApJ...800....9M}.
However,  the role of NSCs 
and their contribution to the BH binary merger rate in the local universe 
 remains quite obscure. 
As discussed in what follows,
 NSCs differ from lower mass GCs in at least three important ways,
each of  these can significantly enhance the BH merger rate and affect the properties
 of the merging binaries in NSCs.

(i) \emph{NSCs retain most of their BHs}.
While natal kicks can easily eject  BHs from GCs,
the natal kick magnitudes are unlikely to be large enough to eject 
a considerable number of BHs from NSCs given the large escape speed
in these latter systems.
Whether dynamically formed BH binaries will merge, and whether the merger
will occur inside  the cluster also depends on the cluster escape speed.  
The low escape speed ($\lesssim 10\rm\ km\ s^{-1}$)  from 
low mass clusters ($M_{\rm cl}\lesssim 10^5\  M_\odot$),  
implies  that most BH binaries are ejected early
after their formation with an orbital 
semi-major axis which is typically too large for 
 GW emission to become efficient and drive the merger of the binary
in one Hubble time.
The vast majority of dynamically formed BH binaries in GCs are also kicked out  before merging,
but they are able to merge in the local universe \citep[e.g.,][]{2015PhRvL.115e1101R}. 
As argued by \citet{2009ApJ...692..917M} given that
 NSCs   have escape speeds that are several times those of globulars, they can retain most of their  BH binaries. Moreover, as we show below, 
 even when accounting for the recoil kick due to anisotropic 
emission of  GW radiation a large fraction of merger products is likely to be retained
inside NSCs. 

(ii) \emph{NSCs contain  young stellar populations.}
The common finding emerging from spectroscopic surveys
 is that NSCs are characterized by  complex star formation histories
  with a mixture of morphological components and different stellar populations spanning a wide range of characteristic ages (from 10 Myr to 10 Gyr) and
  metallicities \citep{2004ApJ...601..319F,2006AJ....132.1074R,2006ApJ...649..692W,2015ApJ...809..143D}. 
  This implies that unlike GCs, NSCs can still form fresh BHs and BH binaries 
  at the present time.
  The presence of significant additional gas not found in old globular clusters could also result in differences in the black hole mass distribution, as well as the dynamics of the underlying black hole population \citep{2013MNRAS.429.2997L,2014MNRAS.441..919L}.

(iii) \emph{NSCs reside at the center of galaxies.}
Therefore, unlike GCs, NSCs are not isolated.
 In time, newly formed star clusters could
  migrate by dynamical friction from the galaxy into the NSC itself
  replenishing   BHs that have been kicked out by three-body processes 
  or by GW recoil kicks. The 
  orbital decay  of massive star clusters through dynamical
  friction  constitutes an additional source 
  which can repopulate the  BH binary population 
 in the nuclei of galaxies \citep{2014ApJ...794..106A}.

 In this paper we study the dynamical formation of BH binary mergers
 in NSCs,   with particular focus on  NSCs which do not host a central massive black hole
 (MBH) {which we define here as BHs having a mass of $\gtrsim 10^6\rm M_\odot$}. 
Our cluster models 
 are based on a semi-analytical approach 
which describes the formation and evolution
 of BH binaries in static cluster models.
 Although necessarily approximated, these models are shown to give reasonable results
 when compared to  recent  Monte Carlo models
 of massive GCs \citep{2016arXiv160300884C,2016arXiv160202444R} and previous 
BH binary merger rate estimates from NSCs \citep{2000ApJ...528L..17P,2009ApJ...692..917M}.
We stress  that although the MBH occupation fraction in NSCs is largely unconstrained observationally,
  it has  been long recognized that
some NSCs do not have MBHs \citep[e.g.,][]{2001Sci...293.1116M,2012AdAst2012E..15N}. 
We note that NSCs   with MBHs are very different, dynamically, than NSCs without.
 If a MBH is present  the velocity dispersion
   keeps growing  towards the MBH, which means that no binary
   will be hard all the way to the center.  
Here we make use of the semi-analytical galaxy formation models
presented in \citet{2015ApJ...812...72A} to predict the occupation fraction of MBHs in NSCs and the NSC initial mass function, which combined with the results of our cluster models  allows an estimate of the aLIGO detection rate and properties of
BH mergers forming in NSCs.
  
 Our results suggest that the BH merger event rate from  NSCs 
 is substantial, with  several tens of events per year  detectable with aLIGO. 
In addition, we propose a new dynamical  pathway to the formation of high mass
BH binary mergers similar to GW150914. This 
merger path is exclusive to 
NSCs and to the most massive GCs.
Due to their large escape speeds,
such  massive clusters can
keep a large fraction of their BH merger remnants while also evolving rapidly enough that the holes can sink back to the central regions where they can form a new binary, which will subsequently harden and merge. We find that  this process can
repeat several times and produce  BH
mergers  of several tens of solar masses
and up to a few hundreds of solar masses,
  without the need  of invoking extremely low
metallicity environments.
 
The paper is organized as follows.
In Sections\ \ref{3rec}  we discuss the processes 
leading to the formation and merger of BH binaries in the
high density cores of GCs and NSCs, focusing on the 
processes that can lead
to the full ejection of BHs. 
In Section\ \ref{sam} we describe our semi-analytical approach and derive the
expected merger rate of BH binaries in NSCs.
In Section\ \ref{disc} we discuss the implications of our results 
including the  aLIGO detection rate and the contribution to the BH merger rate from 
NSCs hosting central MBHs. Finally, we summarize the main results of our study in
Section\ \ref{summ}.

\begin{figure}
\centering
\includegraphics[width=3.1in,angle=270.]{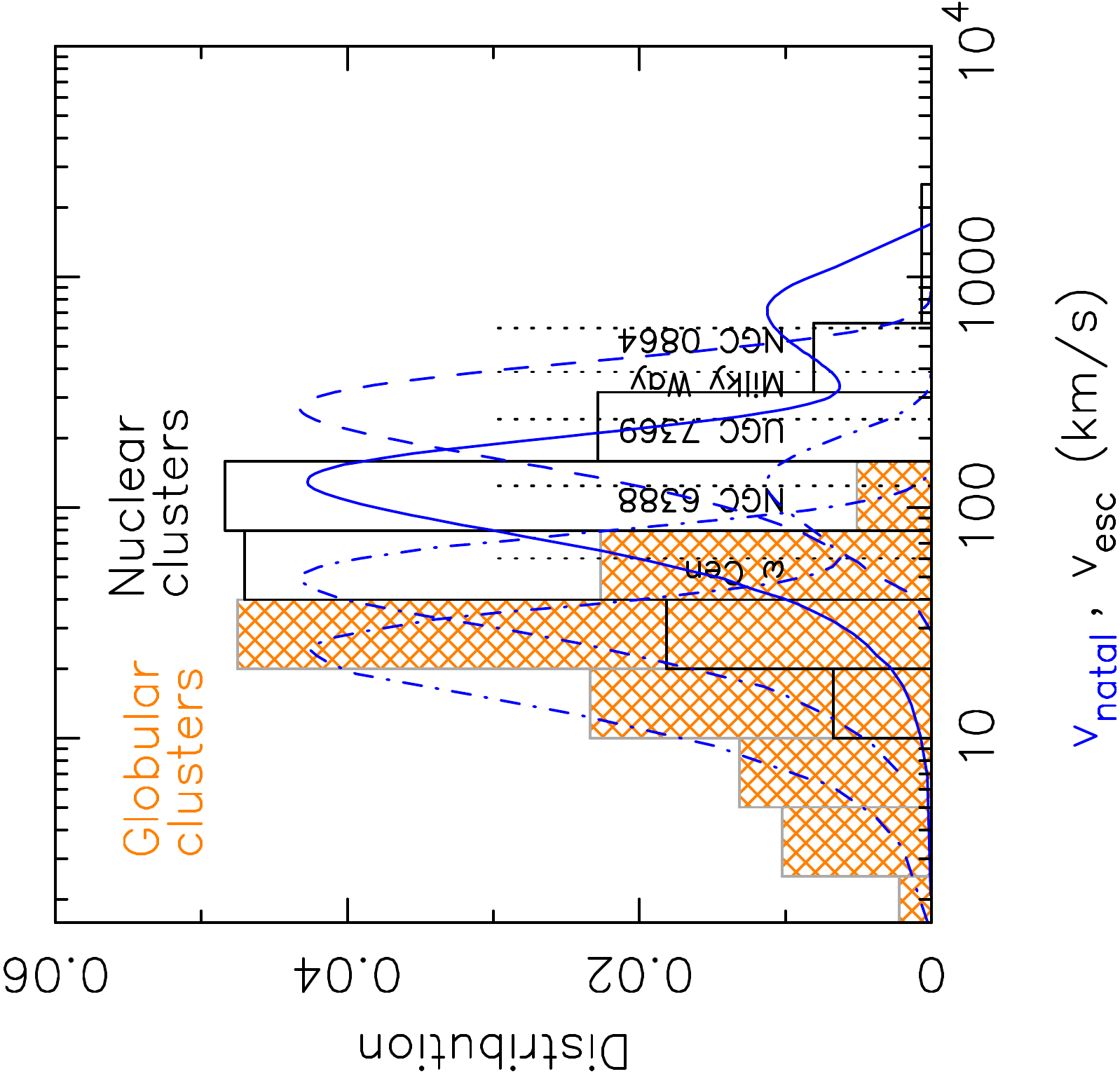}
\caption{Distribution of escape velocities from NSCs and GCs 
  (histograms)   compared  to distributions of
   natal kicks taken from Figure 3 of \citet{2012MNRAS.425.2799R} (blue curves).
   Blue solid and dashed lines correspond to 
distributions that are typically used to model    the kick velocities of neutron stars.
The solid line is the   Arzoumanian distribution \citep{2002ApJ...568..289A}, 
the
   dashed line is the   \citet{1997MNRAS.291..569H}  distribution. 
   The two 
   dot-dashed lines are these two distributions but with kick speeds reduced, assuming that the momentum imparted to the black hole is the same as the momentum imparted to the neutron star. Note that  if BHs receive natal kicks as large as those of neutron stars, most of them will be ejected from
  GCs but not from NSCs.
}\label{Fig0}
\end{figure}

\section{Dynamical evolution of BH binaries in stellar clusters}\label{3rec}

\subsection{Natal kicks}\label{nk0}
 
Due to to asymmetries in the mass ejection 
or in the neutrino flux during core-collapse supernovae
the  black holes might receive appreciable natal kicks which could eject them
from the cluster.
Thus, before we discuss the dynamical processes that can
lead to the formation and merger of BH binaries
in star clusters it is useful to consider 
natal kicks as a phenomenon  that
can fully eject BHs from a star cluster thereby aborting the dynamical formation
channel for BH mergers.

The histograms in Figure\ \ref{Fig0} show the distribution of escape velocities 
from NSCs and GCs.
The escape velocities of GCs are central escape velocities 
 calculated using the photometric data from the catalog by Harris 
 \citep{1996AJ....112.1487H}
 and using 
  single-mass  King models
with a constant mass-to-light ratio $M/L_v = 3$.
The NSCs escape velocities  (from the cluster half-mass radius) were
 computed from the expression \citep[e.g.,][]{2009MNRAS.396.1075G}:
\begin{equation}\label{vesc}
v_{\rm esc}\approx f_c \sqrt{\frac{M_{\rm cl}}{M_\odot}
\frac{\rm pc}{r_{\rm h}}} \ \rm km\ s^{-1} ,
\end{equation}
where $r_{\rm h}$ and $M_{\rm cl}$ are the cluster half-light radius
and mass;
the coefficient $f_{\rm c}$ takes into account the dependence of the
escape velocity on the concentration of the cluster (i.e., $c=\log(r_t/r_c)$, with
$r_t$ and $r_c$ the cluster tidal and core radii).
The cluster radii and masses were taken from the 
sample of late-type galaxies of
\citet{2016MNRAS.457.2122G}.  For more than half of the NSCs in these galaxies
\citet{2014MNRAS.441.3570G} find that a
King profile with a high concentration index, $c=2$, provides the
best fit. This concentration parameter corresponds to
$f_c\approx 0.1$ \citep{1962AJ.....67..471K} -- this latter is the value of
$f_c$ that we adopted in evaluating Eq.\ (\ref{vesc}).
Figure\ \ref{Fig0} 
shows that escape velocities from NSCs are substantially
larger than those from GCs although the two distributions somewhat
overlap near  $M_{\rm cl}\sim 10^6\ M_{\odot}$ where the two type of
systems have similar structural properties \citep[e.g.,][]{2004AJ....127..105B,2006ApJS..165...57C}.

The distributions of natal kicks ($v_{\rm natal}$; blue lines) in Figure\ \ref{Fig0} 
were taken  from Figure 3 of \citet{2012MNRAS.425.2799R}.
These authors consider two different neutron star natal kick distributions. One is 
 the \citet{1997MNRAS.291..569H} distribution, the other is
 the bimodal distribution for neutron star kicks proposed by \citet{2002ApJ...568..289A}
 which has a lower peak at  $\approx 1000\rm\ km\ s^{-1}$ 
  and the higher peak at $\approx 100\rm\ km\ s^{-1}$. 
We also show two modified versions of these distributions (blue dot-dashed lines),
which were obtained by  assuming that the momentum imparted on a 
BH is the same as the momentum given to a neutron star taken from the two
former  distributions. Thus the kick velocities 
are reduced in these latter models by the neutron star to BH mass ratio \citep{2012MNRAS.425.2799R}.

Figure \ref{Fig0} shows that BHs receiving natal kicks as large as $
v_{\rm natal} \gtrsim 50\rm\ km\ s^{-1}$
will escape from GCs before they can dynamically interact, 
which will suppress
the dynamical formation of BH mergers in these systems. 
However,   from Figure \ref{Fig0} we also see  that 
 BHs will be easily retained in NSCs even for natal kicks as large as 
 a few $100\rm\ km\ s^{-1}$. 
 Hence, if BHs receive natal kicks of  
 $\gtrsim 50\rm\ km\ s^{-1}$, we expect that this will greatly reduce the BH  merger
rate from GCs \citep[e.g.,][]{2016arXiv160300884C} as well as that from isolated binary evolution
\citep[e.g.,][]{2016ApJ...819..108B}
 virtually to zero, but it will not significantly 
alter the merger rate of BH binaries
formed dynamically in  NSCs unless the birth kick velocities are 
$\gg 100\rm\ km\ s^{-1}$.
As we will show in Section \ref{bkick} these basic predictions
are in agreement with the results of our cluster models; 
for now we note that the obvious consequence 
of the comparison shown  in  Figure\ \ref{Fig0}
is that the NSC $vs$ GC relative contribution to the BH merger rate 
will depend on the poorly constrained natal kick velocity distribution.
In the following  we assume that at least some BHs are retained 
inside the cluster and consider the  subsequent formation and dynamical evolution
of binary BHs.

\subsection{Mass-segregation}
After few million years from the birth of a star cluster, the most
massive stars explode in supernovae or collapse directly to form BHs.
If the BHs are not ejected by their natal kicks,
being more massive than a typical main-sequence star, they will 
 migrate to the cluster center via  dynamical friction in a process
that is generally referred to as mass segregation.
In the dense environment of the cluster core BHs can efficiently {form}
binaries which will then harden and eventually merge. 
A useful reference time is the two-body relaxation timescale evaluated
at the half-mass radius of the star cluster \citep{Spitzer}:
\begin{eqnarray} \label{th}
t_{\rm rh} \approx 4.2\times 10^9 \left(15\over \ln \Lambda \right) 
 \left(r_{\rm h}\over 4\rm\ pc\right)^{3/2}\left(M_{\rm cl}\over 10^7 \ M_\odot\right)^{1/2} \rm \ yr
\end{eqnarray}
with  $\ln \Lambda$ the Coulomb logarithm.
On a time $t_{\rm rh}$, two-body
gravitational interactions of stars are important in driving the
dynamical evolution of the cluster. 

\begin{figure}
\centering
\includegraphics[width=3.in,angle=270.]{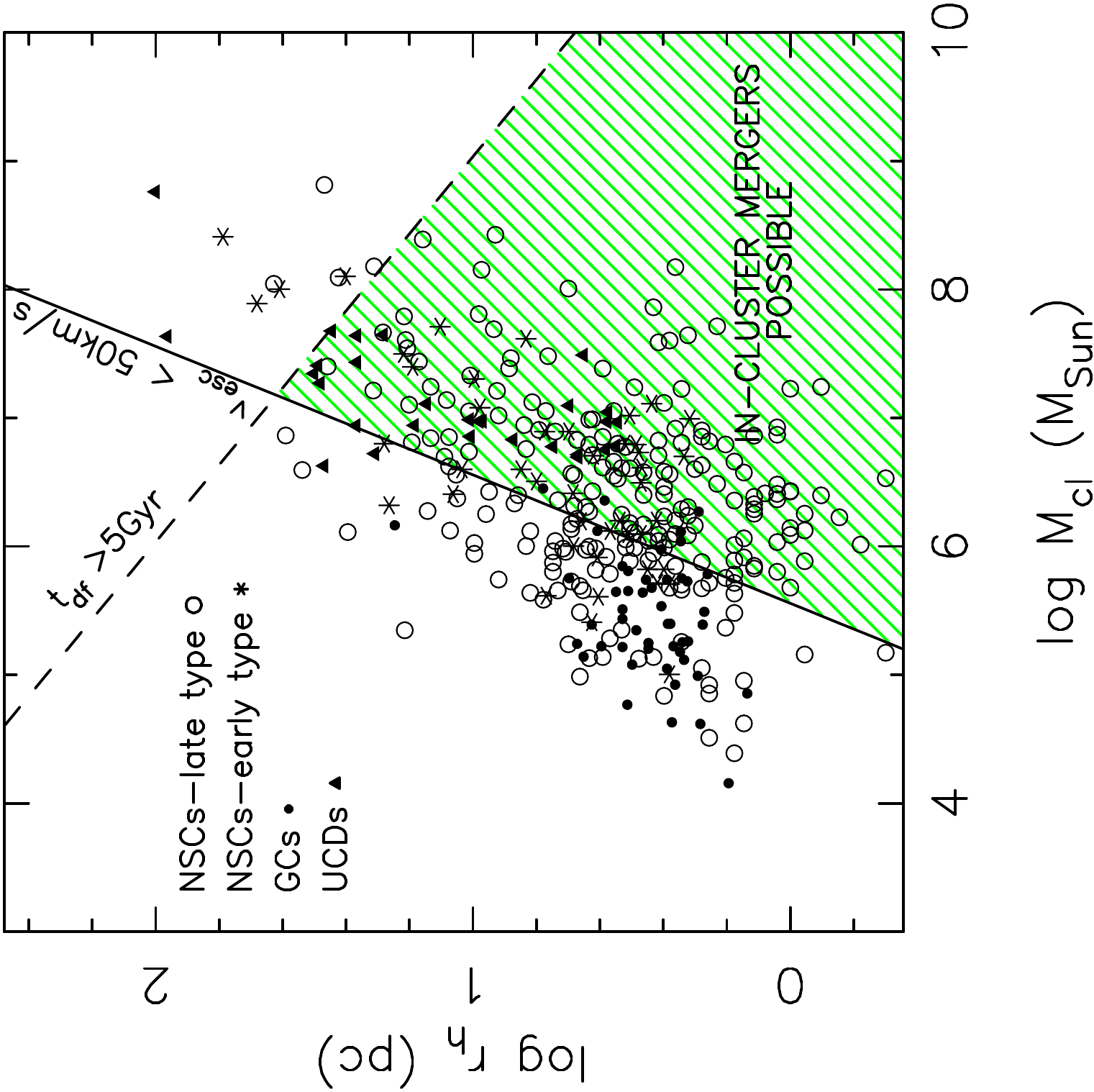}
\caption{
Half mass-radius (or effective radius) against total cluster
  mass for NSCs, GCs, UCDs.
{ 
Data are from \citet{2006ApJS..165...57C} and \citet{2016MNRAS.457.2122G}. 
Systems that lie to the right of the  dashed line
have  $t_{\rm df}>5\rm\ Gyr$. 
Such systems are still
 in the process of forming a BH subsystem. For the majority of the systems we considered,
 including most NSCs and UCDs,
the bulk of the stellar mass BHs are likely to have already
experienced significant mass segregation. 
Systems that lie to the left the solid black line
have  $v_{\rm esc}<50\rm\ km\ s^{-1}$.  This is an indicative value of
cluster escape
velocity  below which  BH binaries will be ejected from the cluster
before merging. In systems that are within the green hatched region
the BHs are more likely to merge while still inside their host cluster.}
}\label{Fig2}
\end{figure}

{ 
While low-mass NSCs have short relaxation times,
for some of the most massive NSCs the  half-mass relaxation time
can exceed the Hubble time. However,
even in the most massive NSCs the  BHs can still segregate at the center on the much shorter dynamical
friction timescale \citep{1943ApJ....97..255C}.}
More precisely, the BHs
will decay to the cluster core on the  timescale \citep[e.g.,][]{BT:87}:
\begin{equation}\label{df}
t_{\rm df}\approx 0.42\times 10^9\left(10\frac{m_\star}{m_\bullet}\right) \left( t_{\rm rh}\over 4.2\times 10^9\rm \ yr \right)\rm \ yr,
\end{equation}
with   $m_\star$ and $m_\bullet$ the mass of a typical cluster star and BH  respectively.
After a time $t_{\rm df} $
the BHs will dominate the densities inside the cluster  core \footnote{Note that Eq.\ (\ref{th}) and Eq.\ (\ref{df}) are strictly valid only for a singular isothermal sphere model.}.

In Figure\ \ref{Fig2} we plot the half-mass radius (or effective
radius) versus the total stellar mass for various types of compact clusters:
NSCs, GCs and Ultra Compact Dwarfs (UCDs). The dashed line delineates
the region below which the dynamical friction timescale becomes
shorter than $\approx 5\rm\ Gyr$
 suggesting  that  the BHs in these systems will sink to the center in
 much less than a Hubble time.
Virtually  all systems we considered  but the most massive NSCs 
and UCDs ($M_{\rm cl}\gtrsim 10^8\ M_{\odot}$)
 can evolve rapidly enough so that the BHs will 
sink to the center where they can
participate in dynamical interactions and swap  into  hard binaries.
The formation of such binaries and their dynamical evolution is
discussed in the following.

\subsection{formation of BH binaries, hardening and mergers}\label{mass-se}
After the BHs segregate to the cluster core, BH binary formation
can efficiently occur through the processes described below.

During core-collapse, if  the densities of BHs become sufficiently  high, BH binaries
can be  assembled through three-body processes in which 
 a binary is formed with the help of a third BH, which carries away the excess
energy needed to  bound the pair \citep{1995MNRAS.272..605L,2015ApJ...800....9M}.
The timescale for three-body binary formation can be written as 
\citep[e.g.,][]{1995MNRAS.272..605L}:
\begin{eqnarray}\label{T3bb}
t_{\rm 3bb} &\approx& 4\times 10^{9} \left(n\over 10^6
{\rm\ pc^{-3}}\right)^{-2} 
\left({\zeta^{-1} {\sigma\over 30 \rm\ km\ s^{-1}}}\right)^{9} \\
&& 
\left( {m_\star\over m_\bullet}
10\right)^{9/2}
\left(  m_\bullet\over 10M_\odot \right)^{-5} \nonumber
\rm \ yr \ ,
\end{eqnarray}
with $n$ the  number  density  of
black holes near the center.
The constant $\zeta\leq 1$ in the previous expression
 parametrizes the departure of the cluster
from equipartition and we have used the relation $ m_\bullet (\zeta\sigma_{\rm BH})^2=
m_\star \sigma^2$ in order
to express $t_{\rm 3bb}$  in terms of the cluster stellar velocity dispersion.

In addition to three-body binary formation, BH binaries
can potentially form through exchange interactions involving primordial stellar binaries
\citep{2009ApJ...692..917M}.
{ Exchange interactions can lead to the efficient formation of 
BH binaries  only if the cluster contains a number
of hard binaries. Thus, this channel might be somewhat 
suppressed in NSCs
-- because of the larger velocity dispersion than in GCs a larger fraction
of binaries will be soft and will be quickly ionized in NSCs.}
 However,
as argued in \citet{2009ApJ...692..917M}, the reduction is not going
to be by a large fraction given that  binaries are typically born
with roughly equal probability per logarithmic interval of 
semi-major axis, $ {\rm d}P/{\rm d}\log (a) = \rm const.$,  in the range $10^{-2}-10^{3}\rm\ AU$ \citep{1991AA...248..485D}.
If, for example, we consider a NSC with velocity dispersion $\approx
30\rm\ km\ s^{-1}$  all binaries with semi-major axis $a\lesssim 1\rm\
AU$ will be hard. If we assume a constant probability per $\log (a)$ for 
$0.01 < a < 10^3\ $AU, then the probability of finding a binary in the
range  of $a = 0.01-1$ AU is substantial, $\approx 40$ percent.
When a BH gets within a couple of semi-major axis lengths of a binary,
the binary will be broken apart and the BH will tend to acquire a companion.
The characteristic timescale on which such exchange interaction occurs
 is $t_{\rm 1-2}=(n\Sigma \sqrt{3}\sigma)^{-1}$,
where $\Sigma=\pi r_p^2\left[1+2Gm_{123}/(\sqrt{3}\sigma
  )^2r_p\right]$  is the interaction cross section for periapsis
distances $\leq r_p\approx 2\rm\ AU$ and $m_{123}$ is the total mass of the
interacting objects.
If the cluster core is dominated by stellar binaries, then 
the timescale for a BH to be capture into a binary is  \citep{2009ApJ...692..917M}:
\begin{eqnarray}\label{12-stars}
t_{\rm 1-2} &\approx& 3\times 10^9 \left( f_b \over 0.01 \right)^{-1}\left(n\over 10^6
{\rm\ pc^{-3}}\right)^{-1}\left({\sigma\over 30 \rm\ km\ s^{-1}}\right) \nonumber \\
 && \left(m_{123}\over 10M_\odot \right)^{-1} \left(a_{\rm hard}\over
  1\rm\ AU\right)^{-1}\rm \ yr \ ,
\end{eqnarray}
where $a_{\rm hard} $ is the typical semi-major axis of hard binaries
and  $f_b$ is the core binary fraction.
By comparing  the previous equation to Eq.\ (\ref{T3bb})
we see that  even under quite standard conditions (but even more so during core-collapse),
 three-body binary formation 
likely dominates the initial dynamical formation of  BH binaries
in NSCs \citep{2015ApJ...800....9M}. 

Two-body binary formation can also occur through gravitational bremsstrahlung
in which two initially unbound BHs become bound after a close encounter 
in which energy is dissipated through gravitational wave radiation. 
If a BH binary is formed in this manner it mergers almost immediately, without further interactions.
However, \citet{1995MNRAS.272..605L} showed
 that for  velocity dispersions $\sigma \lesssim 100\rm\ km\ s^{-1}$
and numbers of BHs $\lesssim 10^3$ expected in the most massive 
star clusters we
study here, the rate of  binary formation from
gravitational bremsstrahlung is much less than that of regular three-body binary formation.
Therefore, for our investigation, we do not account for 
 binary formation though 
gravitational bremsstrahlung, but we caution  that this process could become important
in the most massive NSCs.
In addition, we assume that after BH binaries are formed, binary-single interactions 
dominate over binary-binary interactions, which will be the case
unless the  binary fraction is very high \citep[$\ge 50\%$;][]{2015ApJ...800....9M}.

After BH-binaries are formed they will dominate the
dynamics inside the cluster core.
Assuming the interaction is now between  three BHs each with mass
$10\ M_{\odot}$,
the typical timescale on which a three-body interaction occurs is:
\begin{eqnarray} \label{12}
t_{\rm 1-2} &\approx& 3\times 10^8 \zeta^{-1} \left( f_b \over 0.01 \right)^{-1}
\left(n\over 10^6\ \rm pc^{-3}
\right)^{-1}\left({\sigma\over 30 \rm km\ s^{-1}}\right)\nonumber  \\
&&\left( {m_\star\over m_\bullet}
10\right)^{1/2}  \left(m_{123}\over 30M_\odot \right)^{-1} \left(a_{\rm hard}\over
  1\rm\ AU\right)^{-1}\rm \ yr \ .
\end{eqnarray}
Given that three-body encounters
tend to pair the most massive BHs participating in
the interaction, we expect that
after a time $\lesssim t_{\rm 1-2}$ the most massive BHs in the cluster
will become part of a hard binary.

After a hard binary is formed
it will  tend to harden at a constant rate \citep[e.g.,][]{1996NewA....1...35Q}
\begin{equation}
{{\rm d} a\over {\rm d}t} \Big|_{\rm dyn}=-H \frac{G\rho}{\sigma} a^2\ .
\end{equation}
In this last expression $\rho$ is the local density of stars and BHs,
$H\approx 20$ is the binary hardening rate
and we have assumed all equal mass interlopers.

If after a single interaction with a cluster member of mass $m_\bullet$
the semi-major axis of the binary decreases from $a$
to $a_{\rm fin}$,
then a binary with components of mass $m_1$ and $m_2$ 
will  recoil with a velocity $v_{\rm 2-1}^2=G\mu {m_\bullet\over
  {m_{123}}} \left(1/a_{\rm fin}-1/a\right) \approx 0.2 G\mu {m_\bullet\over
  {m_{123}}} {q_3/a}$, where $\mu=m_1m_2/m_{\rm 12}$, $m_{\rm
  12}=m_1+m_2$, $m_{123}=m_1+m_2+m_\bullet$ and
$q_3=m_\bullet/m_{12}$.
In deriving the previous expression we have assumed that in the interaction
the binding energy of the binary increases by a fraction
$\approx 0.2q_3$ \citep{1996NewA....1...35Q}.
 The previous expressions can
be used to derive the limiting semi-major axis below which a three
body interaction will eject the binary from the system:
\begin{eqnarray}
a_{\rm ej}&= &0.2G\mu  {m_\bullet\over m_{123}} {q_3}/{v_{\rm esc}^2}\\ 
&& = 0.07 
\left(\mu  {m_\bullet\over m_{123}} {q_3}
\frac{1}{M_{\odot}}\right)
\left(\frac{v_{\rm esc}}{50 \rm km\ s^{-1}} \right)^{-2} \rm AU
\nonumber
\ .~~~~~
\end{eqnarray}

The binary keeps hardening at a constant rate until either GW radiation takes
over and drives its merger or it is ejected from the cluster. 
The time evolution of the binary semi-major axis due to GW radiation is
described by the orbit averaged evolution equation \citep{1964PhRv..136.1224P}:
\begin{eqnarray}
{{\rm d} a\over {\rm d} t} \Big|_{\rm GW}&=&
-{64\over 5}\frac{G^3m_1m_2m_{12}}{c^5a^3(1-e^2)^{7/2}} \\
&&\left(1+{73\over24}e^2+\frac{37}{96}e^4\right) \nonumber \ ,
\end{eqnarray}
where $e$ is the binary eccentricity.
The merger time for the two BHs  is:
\begin{equation}\label{tgw}
t_{\rm GW}\approx 2\times10^9 \left({m_1m_2m_{12}\over
    10^3M_{\odot}^3}\right)^{-1} \left(a\over 0.05\rm\ AU
\right)^4(1-e^2)^{7/2}\rm \ yr \ .
\end{equation}
Comparing the above expression  with the expression
for $a_{\rm ej}$  demonstrates an
important point: since the larger the escape velocity from the cluster the smaller
$a_{\rm ej}$, BH binaries that are produced in NSCs  will have
shorter merger time and
are therefore more likely
to merge within  one Hubble time than BH binaries from
lower-mass GCs. 

Let $a_{\rm GW}$ be the semi-major axis at
which GW radiation begins to dominate the energy loss from the
binary. 
A reasonable choice is to set $a_{\rm GW}$ equal to the
semi-major axis at which ${{\rm d} a/ {\rm d}t} |_{\rm dyn}={{\rm d}
  a/ {\rm d}t}
|_{\rm GW}$. Assuming a circular binary, this leads to the relation \citep[e.g.,][]{david-book}:
\begin{eqnarray}\label{agw}
a_{\rm GW}&=&0.05 \left({m_{12}\over
20\ M_{\odot}}\right)^{3/5} \left(q\over{(1+q)^2}\right)^{1/5}\\
&& \left({\sigma \over 30{\rm\ km\ s^{-1}}}\right)^{1/5}
\left( 10^6\ 
  M_{\odot} {\rm\ pc}^{-3}\over \rho \right)^{1/5}
\rm\ AU \nonumber \ ,
\end{eqnarray}
where $q=m_2/m_1\lesssim 1$.
If $a_{\rm GW}>a_{\rm ej}$ merger happens before ejection.
{ By comparing  Eq.\ (\ref{agw}) with the expression
for $a_{\rm ej}$  we see that  
BH binaries that are produced
in NSCs are less likely to be ejected from the cluster.}

The binary will continue to interact with other  cluster members until it reaches a
semi-major axis $a_{\rm crit}=\max(a_{\rm GW},a_{\rm ej})$.
After the binary has
decayed to $a_{\rm crit}$ (where it spends most of its lifetime) 
the timescale between two consecutive interactions 
becomes \citep[e.g.,][]{2004ApJ...616..221G}
\begin{eqnarray} 
t_{\rm 2-1}&\approx& 2\times 10^7\ \zeta^{-1} \left(n\over 10^6 {\rm\ pc^{-3}}
\right)^{-1}\left({\sigma\over 30 \rm\ km\ s^{-1}}\right)\\
&&  \left( {m_\star  \over \ m_\bullet}10
\right)^{1/2} \left(a_{\rm crit}\over
  0.05\rm\ AU\right)^{-1} \left( m_{12}\over 20\ M_{\odot}\right)^{-1}
\rm \ yr \ . \nonumber
\end{eqnarray}
If we assume as before that each interaction removes a fraction
 $0.2q_3$ of the binary binding energy\citep{1996NewA....1...35Q}, 
then the timescale required to decay to $a_{\rm crit}$ from a much
larger separation is of order \citep{2002MNRAS.330..232C}:
\begin{equation}\label{mt}
t_{\rm merge} \approx 5 q_3^{-1} t_{\rm 2-1}.
\end{equation}

During the hard interaction the interloper will recoil at the speed
$v_3=v_{\rm 2-1}/q_3$.  Thus the field BHs will start being ejected when 
$v_3\gtrsim v_{\rm esc}$, at this point the binary 
semi-major axis is $a_3=a_{\rm ej}/q_3^2$. At a fractional hardening 
of $\approx 0.2q_3$ per interaction, { the mass ejected from the cluster
required in order to shrink the binary semi-major axis from $a_3$ to
$a_{\rm crit}$ is approximately:
\begin{equation}\label{3ej}
M_{\rm ej}\approx m_{12}\ln \left({a_3\over a_{\rm crit}}\right),
\end{equation}
so that for low mass clusters $M_{\rm ej}\approx m_{12}\ln \left({1/ q_3^2}\right)$
and $M_{\rm ej}\approx m_{12}\ln \left(a_{\rm ej}/ a_{\rm GW} q_3^2\right)$ for high mass clusters.
Given that $a_{\rm ej}$  decreases with the cluster escape speed, 
the previous equation 
implies that the larger the cluster mass the fewer BH interlopers will be
ejected,  in addition to fewer binaries being ejected;
when
\begin{equation}\label{3ej2}
v_{\rm esc} \gtrsim 120
\left({\mu}  {m_{12}  \over m_{123}} 
\frac{4}{M_{\odot}}\right)^{1/2}\left({{a_{\rm GW}}\over0.05\ {\rm AU}}\right)^{-1/2}
{\rm \ km\ s^{-1}}
\end{equation}
the BH binary will merge without ejecting any of the field BHs.}

 From the condition $ a_{\rm ej}<a_{\rm GW}$,  we derive
the critical cluster escape velocity 
above which binaries will merge before being ejected through
 hard scattering with surrounding stars:
\begin{eqnarray}\label{limit}
\tilde{v}_{\rm esc}&\gtrsim&110
{\sqrt{q}\over 1+q} 
\left(\frac{m_\bullet}{10\ M_{\odot}}\right) \\
&&
\left(\frac{30\ M_\odot}{m_{123}} \right)^{1/2}
\left(a_{\rm GW}\over 0.05\rm\ AU\right)^{-1/2}
  {\rm\ km\ s^{-1}} \ , \nonumber
\end{eqnarray}
so that for  $m_1=m_2=m_\bullet=10\ M_{\odot}$ 
{ we have $\tilde{v}_{\rm esc}\approx 50\rm\ km\ s^{-1}$.
The solid line in Figure\ \ref{Fig2} shows the locus of points where
the escape velocity from the clusters, $v_{\rm esc}(\rm\ km\ s^{-1})\approx
0.1\sqrt{M_{\rm cl}(M_{\odot})/r_{\rm h}(\rm pc)}$ (see Eq.\
(\ref{vesc}) above), is equal to $50\rm\ km\ s^{-1}$.}
The BH binaries forming in clusters lying to the left of this line are likely to
be ejected before merger.
The approximate relation $v_{\rm esc}\approx 2\sqrt{3}\sigma$
 implies that only clusters with velocity dispersion $\sigma \gtrsim 15\rm\ km\
s^{-1}$  will be able to retain their binaries.
In many NSCs and UCDs stellar mass BH binaries
will merge while still inside the cluster, while
most BH mergers in GCs are expected to occur
outside the cluster unless an initially already massive BH
($M\gtrsim 100\ M_{\odot}$) is present in the system\
\citep{2006ApJ...640..156G}.
This result is consistent with Monte Carlo simulations of
GC models where the vast majority of BH binary assembled dynamically through 
$N$-body interactions are found to merge after escaping from their 
host systems \citep[e.g.,][]{2011MNRAS.416..133D}.

Figure \ref{Fig2} shows that 
in  many NSCs and UCDs, the BH merger remnants are likely to be retained so they
might  { form}  new BH-binaries, that will subsequently harden and
merge. It is therefore possible that BHs in these massive star
clusters will undergo a number of repeated mergers and grow considerably. 

In addition to the recoil kick due to three-body interactions, 
 as two compact objects merge, asymmetric emission of gravitational radiation
will also induce  a  recoil velocity  which
can eject the merger product from the system. 
In the next section we discuss this additional
effect.

\begin{figure*}
\centering
\includegraphics[width=3.in,angle=270.]{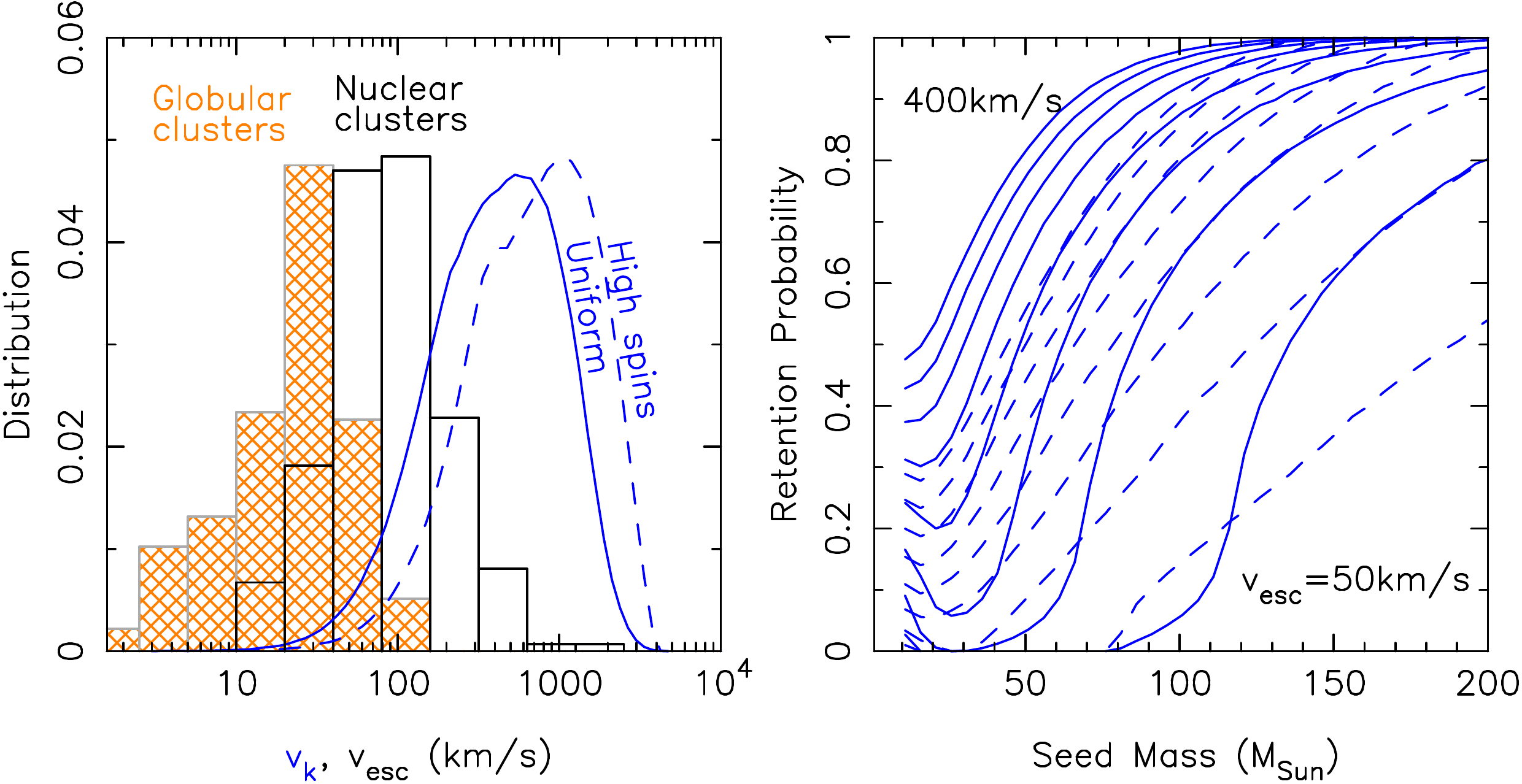}
\caption{Left panel: distribution of escape velocities from NSCs and GCs 
  (histograms)   compared  to the distributions of 
GW kick velocities of merging BHs 
(blue curves).
The solid blue line corresponds to a model in which the spin
magnitude was chosen randomly in the range $\chi= [0, 1)$; the  dashed blue  line
corresponds to a high-spin model in which $\chi=0.9$.
Right panel: probability of remaining inside the cluster as a
function of initial  mass of the dominant BH and for different values
of the cluster escape velocity. Here we have assumed
that  the mass of the secondary BH is $10\  M_\odot$. Solid line is for
the uniform spin model; dashed line is for the high-spin model.
These plots show that the recoil velocity imparted by the anisotropic emission of GW
radiation will lead to the ejection of most BH merger remnants
formed inside GCs, while in NSCs  a fraction of BHs will be retained.
For this reason BHs of large mass 
can naturally grow inside NSCs through repeated accretion of lower
mass BHs. Thus, NSCs are a likely  host environment for  the
highest-mass
BH mergers that are potentially detectable by aLIGO.
}\label{Fig1}
\end{figure*}

\subsection{Gravitational wave recoil}\label{gwr}
The GW recoil velocity of a merged BH depends on the mass ratio and spins of the progenitor BHs.
Hence, in order to make predictions about  the distribution of recoil velocities for
dynamically formed BHs in star clusters
we first define the pre-merger BH spin and
mass distributions.

The mass distribution of BH binaries is quite uncertain.
Here, we use the BH mass distribution
of the dynamically formed  merging black hole
binaries from the Monte Carlo models of massive star clusters
presented in \citet{2016arXiv160300884C}.  These distributions 
contain no BH binary with mass ratio less than $q\approx 0.5$. 
This is expected
given that dynamical encounters in star clusters tend to pair and
eject tight BH binaries with similar mass components --
binaries in dense stellar environments are prone to exchange
components, preferentially  ejecting lighter partners
 in favor of more massive companions \citep{1993Natur.364..423S}.

The distribution of BH spins is also very uncertain.
If the BH inspiral is driven predominantly by random gravitational interactions with other
BHs and stars we might expect the spin orientations to be close to
random. We note that if there is significant coherent gas accretion, the spins might
align in a way that might lead to low recoil kicks
\citep[e.g.,][]{2007ApJ...661L.147B}. But
 stellar-mass BHs are unlikely  to accrete
enough mass from the interstellar medium for this process to be effective
\citep{2009ApJ...692..917M}. 
For these reasons, in our computations the misalignment angle of each
BH is chosen at random in cos$(\theta)$.

For the spin magnitudes of the BHs we consider two choices.
The blue solid line in the left panel of Figure \ref{Fig1} corresponds to a ``uniform'' model in which
 the initial spin magnitudes are drawn uniformly from 
 the range ${\chi}=[0, 1)$,
 where $\vec{\chi}$ is the dimensionless spin of the BH ($\vec{\chi}=\vec{S}/m_\bullet^2$, where
 $\vec{S}$ is the spin angular momentum in units of $m^2$).
The blue dashed line corresponds to one additional ``high-spin'' model
in which  the  spin magnitude is set to a fixed value, $\chi=0.9$.  
We note here that our spin magnitude distributions differ for example from those
of \citet{2016arXiv160202809O} who adopted low-spinning BHs,
leading to low merger kick velocities.
Our choice is motivated by observations: typical estimates of stellar-mass BH spins suggest  high
values, $\chi > 0.5$,  in many cases \citep[for a review see][]{2015PhR...548....1M}. 
In addition, equal-mass non-spinning binaries produce a rotating
(Kerr) BH with final spin magnitude
$\chi \approx 0.69$ \citep{2016arXiv160501938H}, so 
that  BHs undergoing more than one merger
inside the cluster will have a finite spin magnitude 
\citep{2007PhRvD..76f4034B}.
However, we note that configurations leading to rapidly spinning BHs are  rare.
The dimensionless spin magnitude tends to decrease
for a BH that engages in a series of mergers, if the lighter
BHs with which it mergers have a constant mass \citep{2002ApJ...581..438M,BH:03}. 
This will keep the growing BH safely in the cluster after the first
few mergers: not only does the mass ratio get
farther from unity, which decreases the kick, but the spin of the
more massive black hole drops as well.

After the pre-merger BH spin and mass distributions have been
defined, we compute the recoil kick velocity from the following
fitting formula based on the results from numerical relativity simulations of \citet{lousto+12}:
\begin{equation}
{\vec{v}_{\rm k}} = v_{\rm m}  {{\hat e}_{\perp,1}}+ v_{\perp} ({\rm cos} \,\xi \, {{\hat e}_{\perp,1}} + {\rm sin} \,\xi \, {{\hat e}_{\perp,2}}) + v_{\parallel} {{\hat e}_{\parallel}},
\label{eqn:kick}
\end{equation}
\begin{equation}
v_{\rm m} = A \eta^2 \sqrt{1 - 4\eta} \,(1 + B \eta), 
\end{equation}
\begin{equation}
v_{\perp} = {H \eta^2 \over (1 + q )} (\chi_{2\parallel} - q \chi_{1\parallel}),
\end{equation}
\begin{eqnarray}
v_{\parallel} = {16 \eta^2 \over (1+ q)} \left [ V_{1,1} + V_{\rm A} \tilde S_{\parallel} + V_{\rm B} \tilde S_{\parallel}^2 + V_{\rm C}\tilde S_{\parallel}^3 \right ] \times \nonumber
\\ 
|\, {\vec{\chi}_{2\perp}} - q {\vec{\chi}_{1\perp}} | \, {\rm cos}(\phi_{\Delta} - \phi_1),
\end{eqnarray}
where 
$\eta \equiv q/(1+q)^2$ is the symmetric mass ratio; $\perp$ and
$\parallel$ refer to vector components perpendicular and parallel to
the orbital angular momentum, respectively,  ${{\hat
    e}_{\perp,1}}$ and ${{\hat e}_{\perp,2}}$ are orthogonal
unit vectors in the orbital plane, and
${\vec{\tilde S}} \equiv 2({\vec{\chi}_2} + q^2 {
  \vec{\chi}_1})/(1+q)^2$.
The values of $A = 1.2\times 10^4$
\kms, $B = -0.93$, $H = 6.9\times10^3$\ \kms, and $\xi = 145^{\circ}$
are  from \citet{g07} and \citet{2008PhRvD..77d4028L}, and $V_{1,1} =
3678$\ \kms, $V_{\rm A} = 2481$\ \kms, $V_{\rm B} = 1793$\ \kms,
and $V_{\rm C} = 1507$\kms are taken from  \citet{lousto+12}.
The angle $\phi_{\Delta}$ is that  between the
in-plane component ${\vec{\Delta}_{\perp}}$ of the vector ${
  \vec{\Delta}} \equiv M^2({\vec{\chi_2}} - q {\vec{\chi_1}})/(1+q)$ and the
infall direction at merger.  We take the phase angle $\phi_1$ of the binary to be random.

The histograms  in the left panel of Figure \ref{Fig1} show the escape
velocities from NSCs and GCs computed as described in Section \ref{nk0}.
The blue curves show the recoil
velocity distributions for our   models computed using 
Eq.\ (\ref{eqn:kick}).
The recoil velocity distribution in the uniform spin model 
is peaked at $v_{\rm k}\approx 500$\ \kms, while the high-spin model
produces significantly larger kicks with typical velocities $v_{\rm k}\approx 1000$\ \kms.
Note however that in both models there is a substantial fraction of systems
that are accelerated with velocities $\lesssim 100$\ \kms. 

The left panel of Figure \ref{Fig1} suggests that 
 only the most massive GCs
  have a finite probability of retaining a BH merger remnant
formed inside the cluster. Considering also
that  BH binaries in GCs are likely to be flung before 
merger due to three-body encounters, we conclude
that  the retention probability of BH merger remnants
in GCs is small.
The left panel of Figure \ref{Fig1} shows instead  that
the escape velocities of many  NSCs are high enough that a substantial
number of mergers are expected to be retained inside these systems.

In the right panel of Figure \ref{Fig1} we compute the probability of
remaining in the cluster for our spin distributions
as a function of the initial BH mass and assuming that
the secondary BH mass is $10\ M_{\odot}$.  For escape velocities
$\lesssim 50\rm\ km\ s^{-1}$ (typical of massive GCs) the probability of
remaining inside the cluster after a merger is essentially zero,
unless the cluster contains initially a BH seed of mass $\gtrsim
100\ M_{\odot}$.  For escape velocities $200\rm\ km\ s^{-1}$,
which are more typical of NSCs, the probability of retaining a BH merger
remnant  of initial mass $50\  M_\odot$ is approximately $0.5$ or $0.3$
depending on the assumed spin distribution.
This makes NSCs  excellent candidates
for producing massive BH mergers that are potentially 
observable by aLIGO, because they can retain their BHs  
while also evolving rapidly enough that
the BHs can sink back to the center and dynamically form new
binaries which will subsequently merge. This merger channel is 
expected to occur quite naturally in  massive  stellar clusters such as NSCs and
UCDs, while it is unlikely to happen in lower mass systems such as
open clusters and GCs.

In the next section we present a semi-analytical model that we use in order to
make predictions about the mass distribution and rates of BH binary mergers forming in NSCs.

\begin{figure*}
\centering
\includegraphics[width=7.in,angle=0.]{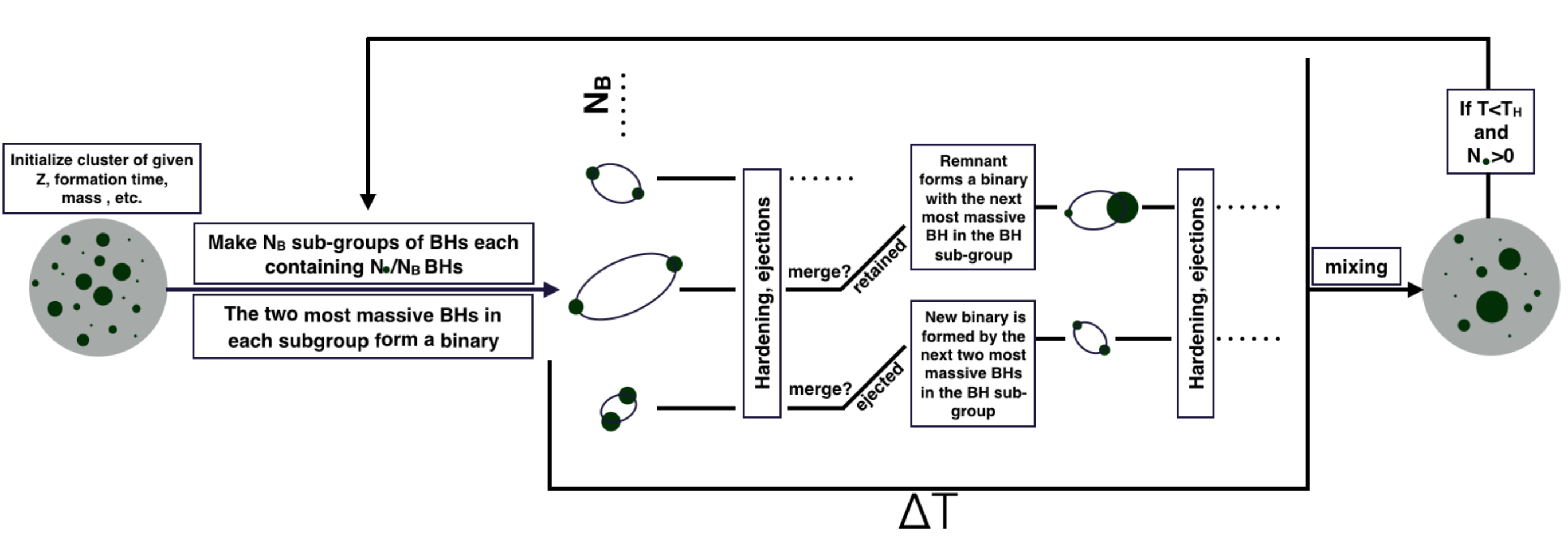}
\caption{
Schematic diagram illustrating our semi-analytical algorithm
to model the evolution of BH binaries in stellar clusters.
After we initialize the star cluster, we divide the BH population
in sub-groups each containing an equal number of BHs. 
Each sub-group contains a BH binary whose components are 
always the two BHs which are currently the two most massive in the sub-group.
The binary is evolved for a time interval $\Delta t$; if the binary merges or is ejected 
from the cluster a new binary is formed and evolved. After a time interval $\Delta t$
all BHs are mixed back together and the procedure repeated until either all BHs have been ejected
from the cluster or the integration time becomes longer then the Hubble time ($T_{\rm H}$).
}\label{dgrm}
\end{figure*}

\begin{figure}
\centering
\includegraphics[width=2.5in,angle=270.]{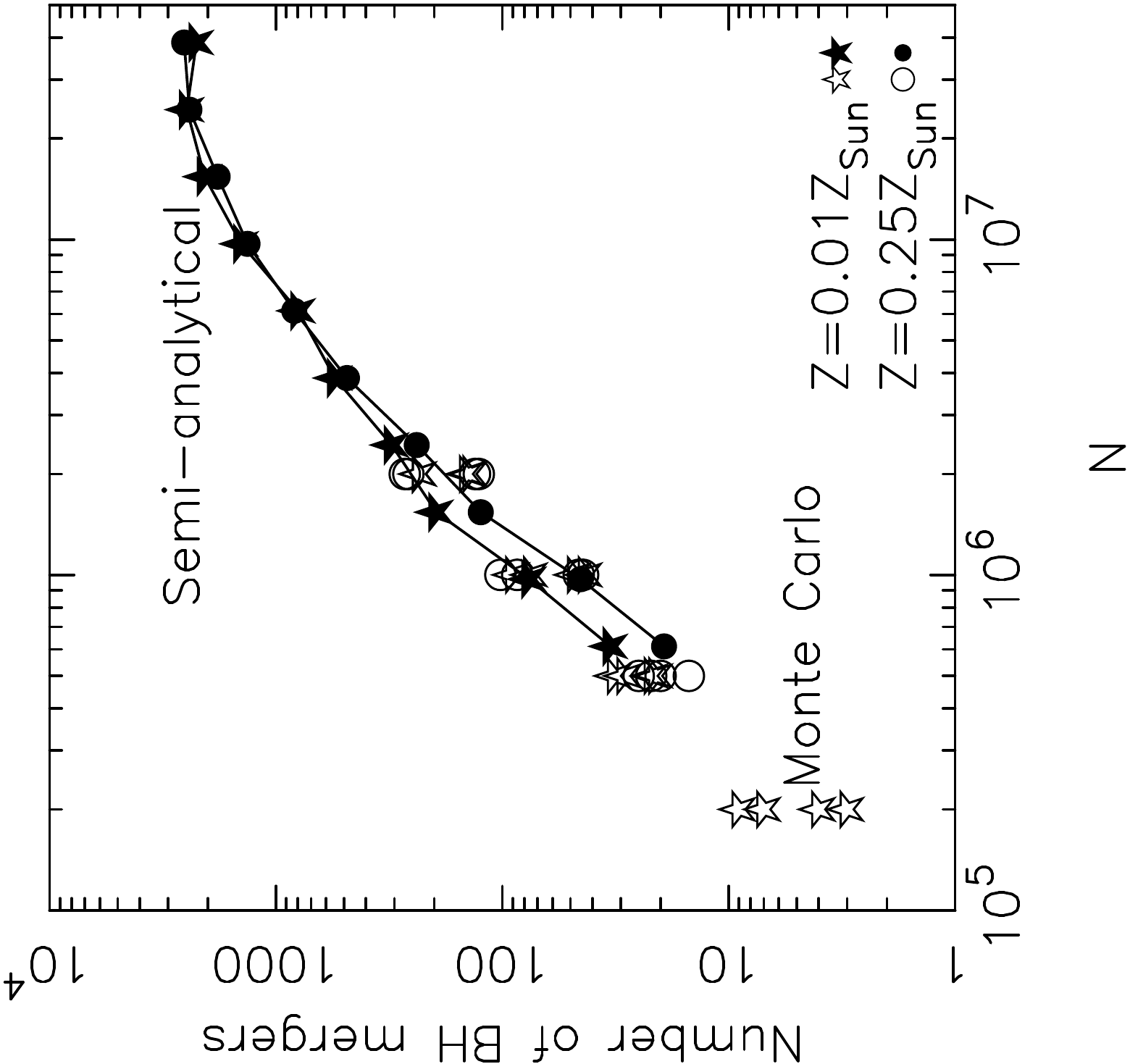}
\includegraphics[width=2.5in,angle=270.]{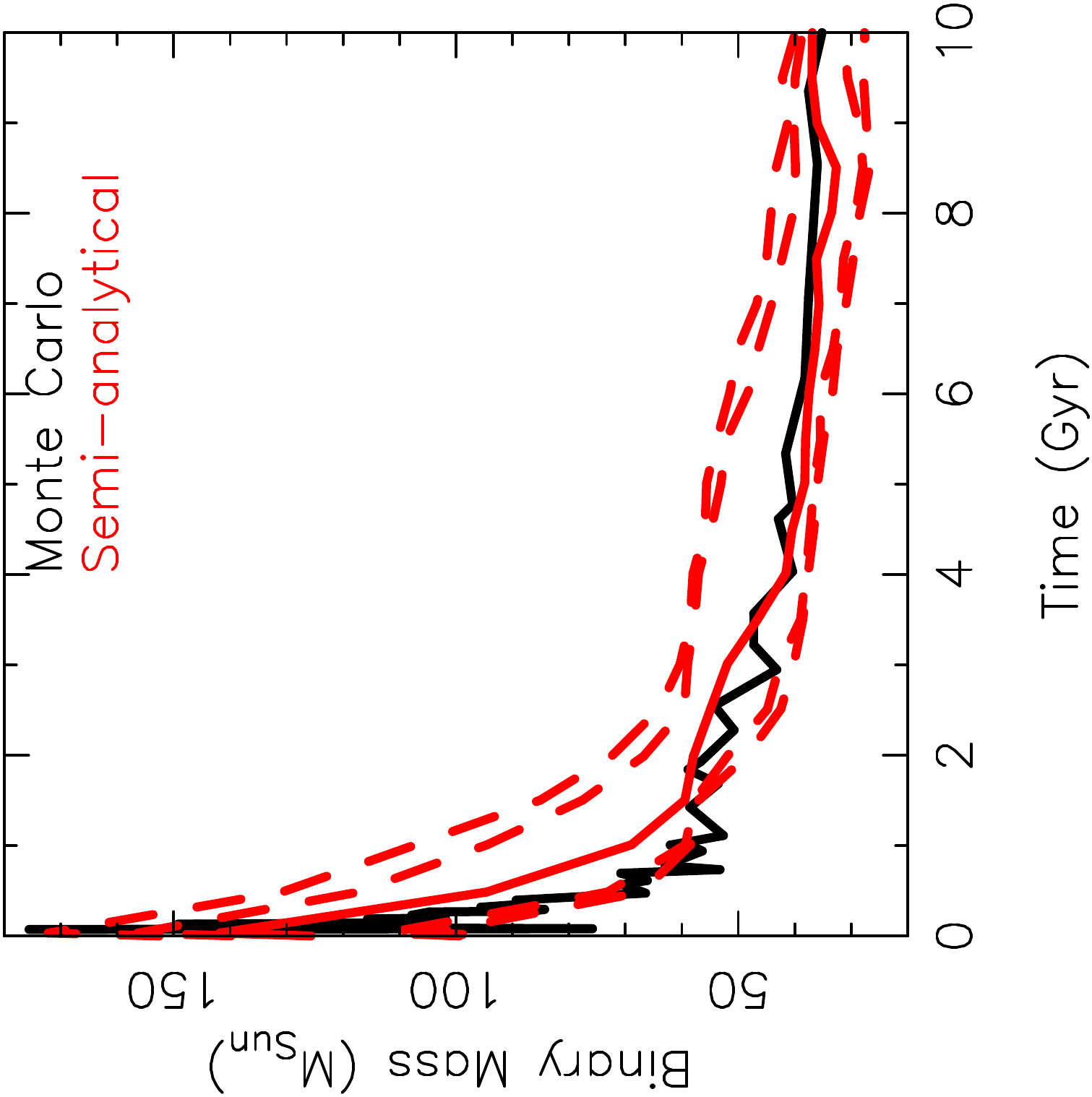}
\caption{{ Comparison between the results of our semi-analytical model 
with the results of  Monte Carlo models
\citep{2016arXiv160202444R}.
Upper panel shows the median number of merging BH binaries as a
function of the total number of stars in the cluster. Open symbols are from the 
Monte Carlo simulations of \citet{2016arXiv160202444R}. Filled symbols
are  from our simplified semi-analytical approach.
Lower panel gives the median mass of ejected BH-binaries 
as a function of time of ejection from 
 a GC Monte Carlo model of \citet{2016arXiv160202444R} (black curve)
and the average mass of ejected BH-binaries  
from ten semi-analytical  
models having similar structural properties (red curve).
Dashed curves give the region containing $70\%$ and $90\%$ of the
ejected systems in these models. Solid curve is the median of the mass distribution.}
}\label{Fig3}
\end{figure}

\section{semi-analytical modeling}\label{sam}
As argued above the dynamical evolution of NSCs 
is of great interest as these systems could represent a 
important source of inspiraling BHs
detectable by aLIGO. Yet a good understanding of
the  dynamical evolution of massive clusters ($M_{\rm cl}>>10^6\ M_{\odot}$)
and their  implications for aLIGO detections  is still elusive. The main difficulty is
the large number of particles comprising these systems which makes 
their treatment extremely challenging even for approximate Monte Carlo methods.
 Here we adopt a semi-analytical approach
which allows us to make predictions about the expected rate
and properties of inspiraling  BH binaries forming in NSCs.

\subsection{Simplified approach}
First, we define the structural properties of our star clusters.
We assign a total stellar mass $M_{\rm cl}$ to the cluster.
For $M_{\rm cl}\leq 5\times 10^6\  M_\odot$  the half-mass radius
is independent on the cluster mass and it is set to $r_h=3\rm\ pc$.
In the NSC mass regime, $M_{\rm cl}> 5\times 10^6\  M_\odot$, we adopt 
the fitted relation to the NSCs in late type galaxies from
\citet{2016MNRAS.457.2122G}:
$\log(r_h/c1)=\alpha\log(M_{\rm cl}/c2)+\beta$, with $\alpha=0.321$,
$\beta=-0.011$, $c1=3.31\rm\ pc$ and $c2=3.6\times10^6\rm\ M_{\odot}$.
{ While sampling from  the adopted  distributions 
we also accounted for the scatter of the observed relations.}
The escape velocity from the cluster is then computed using the approximate
Eq.\ (\ref{vesc}) above;  the cluster velocity dispersion is $\sigma=v_{\rm
  esc}/(2\sqrt{3})$.
The central number density of stars was computed as
$n=4\times 10^6(\sigma/100\rm\ km\ s^{-1})^2\rm\ pc^{-3}$. 
This latter expression gives a central number density of stars for a
Milky Way like NSC of $4\times 10^6 \rm\ pc^{-3}$ 
and $\approx 10^5 \rm\ pc^{-3}$  for a
$10^6\ M_{\odot}$ GC -- this is consistent with observed 
values \citep{1996AJ....112.1487H,2010ApJ...718..739M}.

Next we define the initial mass distribution and number density of  BHs in our cluster models.
We take the mass distribution of single BHs from Figure\
6 of \citet{2016arXiv160202444R}. These authors used the stellar
evolution 
code BSE \citep{2002MNRAS.329..897H,2007ApJ...665..707H}
improved with the stellar remnant prescription from \citet{kh09} and \citet{2010ApJ...719..915C}. 
Our models adopt the update  prescriptions for stellar winds and
supernova fallback,  in order to replicate the BH mass distribution of
\citet{2013ApJ...779...72D} and \citet{2010ApJ...715L.138B}. 
We consider two values of metallicity, $Z=0.01\ Z_\odot$
 and  $Z=0.25\ Z_\odot$, defined below as low metallicity and 
 high metallicity models. 
 
{
In our calculations we assume that all clusters 
formed $12\rm\ Gyr$ ago regardless of their mass. While this is a good
approximation for GCs, NSCs are known to have complex star formation histories, including 
recent episodes of star formation. We neglect such complication in the following, noting 
that the bulk of the stellar population in NSCs is also likely to be  in old stars formed 
many Gyrs ago \citep[e.g.,][]{2011ApJ...741..108P}.
}

Initially, our cluster models have a total mass in BHs that is
$M_\bullet=0.01M_{\rm cl}$.
This is the typical mass fraction in BHs expected for 
standard initial mass functions
\citep[e.g.,][]{2006ApJ...645L.133H}. 
The total number of BHs is therefore $N_\bullet\approx
M_\bullet/\langle m_\bullet\rangle$,
with $\langle m_\bullet\rangle$  the average BH mass in our models.
Then we consider natal kicks. 
For each BH in our fiducial model
we compute a natal kick velocity from a Maxwellian given by 
$\sigma_{\rm natal}=265\rm\ km\ s^{-1}$ as commonly done
for neutron stars \citep{2005MNRAS.360..974H}, and assume that the  natal velocity 
of a BH of mass $m_\bullet$ is lowered
by the factor $1.4\ M_\odot/m_\bullet$.
 For any sufficiently massive BH progenitor ($>40\ M_\odot$), the 
fallback completely damps any natal kick, and the BH is retained in
the cluster \citep{2001ApJ...554..548F}. 
BHs that receive a kick with velocity larger than the escape velocity
from the cluster are removed from our models. However, since we only consider massive
clusters with large escape velocities, a large fraction of BHs in our models is retained after
experiencing a natal kick. This makes our conclusions less sensitive
to the prescription we used for natal kicks, provided that the
real kick magnitudes are not much larger than what we have adopted here.
We discuss in more details the effect  of varying the natal kick magnitudes  below in Section \ref{bkick}.

We assume that after a time $t_{\rm df} (\langle m_\bullet\rangle)$ the BHs have segregated to
the cluster center.
After this time, due to the high  densities in the core, BH binaries will 
efficiently form though 
$3$-body binary formation \citep{2015ApJ...800....9M}
and possibly through exchange interactions with stellar
binaries  \citep{2009ApJ...692..917M}.
Therefore we assume that after a time
$t_{\rm df} (\langle m_\bullet\rangle)$, a 
fraction $f_{\rm bin}=0.01$ of the BHs end up in hard
BH binaries. Although this fraction is quite uncertain,
the value we adopted is  typical for
Monte Carlo models of massive star clusters with low binary fraction \citep[e.g., see Fig. 6
in][]{2015ApJ...800....9M}.  

After the BHs have segregated to the center and we have assigned a
fraction of them to be in BH binaries we follow the evolution, ejection
and formation of new binaries adopting the scheme described in 
what follows.

{
We divide the BH cluster in $N_{\rm B}=f_{\rm bin}N_\bullet/2$ sub-groups, each
containing the same number of BHs. We 
find the two most massive BHs in each sub-group
and assume that after a time $t_{1-2}$ they form a binary. 
Thus, we assume that each
sub-group always contains one binary and that this binary is always
composed of the two most massive BHs in the sub-group.

While  each binary is assumed to evolve in the gravitational potential of the entire cluster
the adopted numerical scheme  allows us to simulate a scenario in which 
the number of BH binaries in the cluster is approximately constant
with time. This, besides  allowing us to greatly simplify
our approach,    appears
 to be reasonable when compared to the results of  Monte Carlo
models of massive clusters \citep[]{2015ApJ...800....9M,2016arXiv160300884C}.
Moreover, we assume that  the BH binaries
 are always composed of the two most massive black holes
in each sub-group
because during exchange encounters lighter partners are more likely to be ejected.
This favors the formation of high mass binaries with similar mass components
 \citep[e.g.,][]{2016arXiv160300884C}.
In our calculation we conservatively 
assume that the interactions occur between BHs and stellar binaries
so that the timescale for binary formation is the longer 
timescale given  by  Eq. (\ref{12-stars}).
The binary fraction in evaluating $t_{1-2}$  was computed taking a primordial 
binary fraction of $0.2$  and lowering this fraction 
by the number of soft binaries for a constant probability in 
$\log(a)$ \citep{1991AA...248..485D}.
}

Any binary  forms with an initial semi-major axis
$a_{\rm hard}=1/(\sigma/30\rm\ km\ s^{-1})^2 \rm\ AU$.
Given the cluster velocity dispersion, its density and
the mass of the binary,  we compute (i) the semi-major axis, $a_{\rm GW}$,
below which GW radiation will start to dominate  
(Eq.\ [\ref{agw}]), and (ii) 
the semi-major axis, $a_{\rm ej}$, at which the binary will be
ejected as a consequence of  three
body scatterings.
 If $a_{\rm ej}>a_{\rm GW}$ the binary will merge
outside the cluster and will be ejected with a
semi-major axis $\approx a_{\rm ej}$; in this case we evaluate the
timescale from the formation of the binary to its ejection, $t_{\rm ej}$,
using Eq.\ (\ref{mt}) so that the lifetime of the binary is
 $T=t_{\rm ej}+t_{\rm GW}(a=a_{\rm ej})$.
 If $a_{\rm ej}<a_{\rm GW}$ the binary will merge
inside the cluster; in this latter case the total lifetime of the binary 
is $T=t_{\rm merge}+t_{\rm GW}(a=a_{\rm GW})$.
In the previous expressions the GW merger timescale, $t_{\rm GW}$,
was computed by sampling the binary eccentricity  from a thermal distribution $N\propto
e^2$.

If the   BH binary is ejected  from the cluster, then 
after a time $t_{1-2}$ we form a new binary and, as before, we take 
its components to be the next two most massive BHs in the sub-group.
Then the hardening timescale of the binary is evaluated as before and 
it is determined whether the new binary will merge inside the cluster,
and, if it does, whether it will be retained inside the cluster after merging.

If the binary merges inside the cluster (i.e., $a_{\rm ej}<a_{\rm GW}$) we assign the two progenitor
BHs a spin magnitude and orientation from the spin models
described in Section\ \ref{gwr} and compute the 
GW recoil speed through Eq.\ (\ref{eqn:kick}). 
In order to account for the recoil kick due to the interaction with a third object
we compute a total kick velocity as $v_{\rm tot}=\sqrt{v^2_{\rm k}+v^2_{2-1}}$,
with $v_{2-1}$ computed as in Section\ \ref{mass-se} 
(note that $v_{\rm tot}\approx v_{\rm k}$ typically).
If $v_{\rm tot}>v_{\rm esc}$ the BH
merger remnant is ejected from the cluster, otherwise it is retained.

If the BH merger remnant is ultimately retained inside the cluster it will have
another chance of interacting with new binaries and experience
additional mergers. In this case, we place the BH remnant at a distance 
$r_{\rm h}(v_{\rm tot}/v_{\rm esc})^2$ from the center and evaluate the
dynamical friction timescale for the BH to reach the cluster core through Eq.\
(\ref{df}). 
If $t_{\rm df}$ is greater than $10\rm\ Gyr$  
the BH is removed from the computation,
 otherwise after a time $t_{1-2}$ 
the BH forms a new binary with the next  most massive BH in the sub-group.
Then, the hardening timescale of the new binary is evaluated as before and 
it is determined whether the new binary will merge inside the cluster,
and if it does whether it will be ejected from the cluster after the
recoil due to anisotropic emission of GW radiation.

{ As the binary  hardens we calculate the number of
field BHs that are ejected through three-body encounters 
as $N_{\rm ej}=M_{\rm ej}/\langle m_\bullet\rangle$ where 
$M_{\rm ej}$ is given by  Eq.\ (\ref{3ej}). If the number of ejected BHs becomes larger
than the total initial number of BHs  we stop the
integration.}

The previous steps (i.e., binary formation, hardening, and ejection/merger)
are repeated  for each sub-group for
a time-step $\Delta t$. After an interval of time $\Delta t$ the
remaining BHs in all the sub-group
are mixed  together and the procedure described above is repeated
recursively until either all BHs have been ejected from the cluster or the total
integration time exceeds the Hubble time. 
The mixing of the sub-groups every $\Delta t$  allows us
to avoid suppressing exchange  interactions between massive BHs
that might  grow in different sub-groups. In what follows we set
$\Delta t=1.5\times 10^9\rm \ yr$, but found that 
values in the range  $\Delta t=1-3\times 10^9\rm \ yr$ all
produced similar results.

The main steps of our semi-analytical algorithm  are 
also schematically illustrated in Figure\ \ref{dgrm}.
We note that our prescriptions are oversimplified in many ways
and that more accurate Monte-Carlo simulations will be needed in order to confirm
our results. 
One basic simplifying  assumption  is  that the 
cluster structural properties (e.g.,  central density, half-mass
radius) remain constant in time.  We believe that this assumption is also justified in many
cases, and especially in very massive clusters where the relaxation
timescale is
longer. For example, Monte Carlo simulations of moderately massive GCs
find that $r_{\rm h}$ increases with time, but often only by a
factor $\lesssim 3$ throughout the cluster
evolution \citep{2016arXiv160300884C}.  
Additionally, in our models we assume
that the binary-single interactions rate is
always dominant with respect to that of binary-binary 
interactions. This latter assumption is also reasonable, unless the
cluster has a very large initial binary 
fraction \citep[$\gtrsim 0.5$;][]{2015ApJ...800....9M}.
Finally, we note that we do not follow 
the evolution of the BH spins through consecutive mergers but assume
that the spins are always drawn from the assumed distributions.

\begin{figure*} 
\centering
\includegraphics[width=5.in,angle=270.]{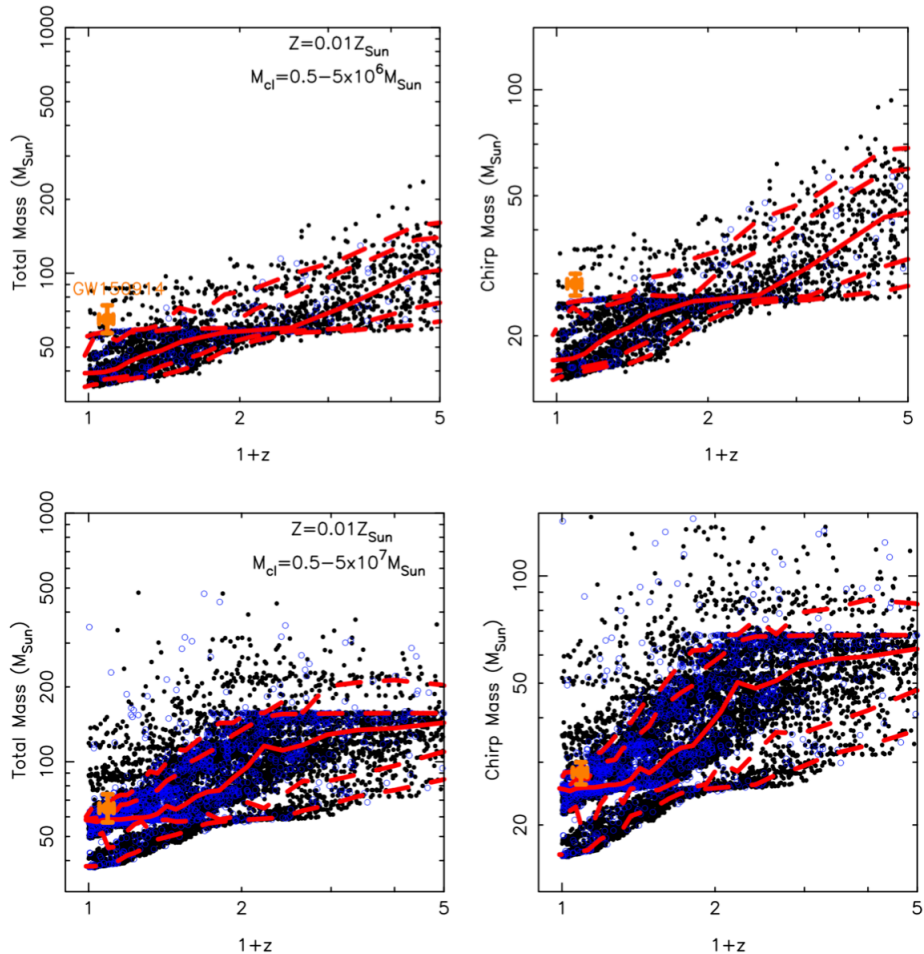}
\caption{{ Mass of merging BH binaries for a range of cluster
  masses which could represent typical GCs (upper panels) or NSCs
  (lower panels) as a function of redshift. Evolution proceeds from right to left. 
     We assume here that 
all clusters formed 12 Gyr ago.}
 The uniform spin model
  described in Section\ \ref{gwr} was adopted. Open blue circles  are systems the are retained
  inside the cluster after merging. Note how almost all mergers 
occurring inside low mass clusters are promptly ejected, while for
$M_{\rm cl}=0.5-5\times 10^7M_{\odot}$ many of the inspiraling BHs are
expected to be retained inside the cluster.
  }\label{Fig4} 
\end{figure*}

\begin{figure*} 
\centering
\includegraphics[width=5.in,angle=90.]{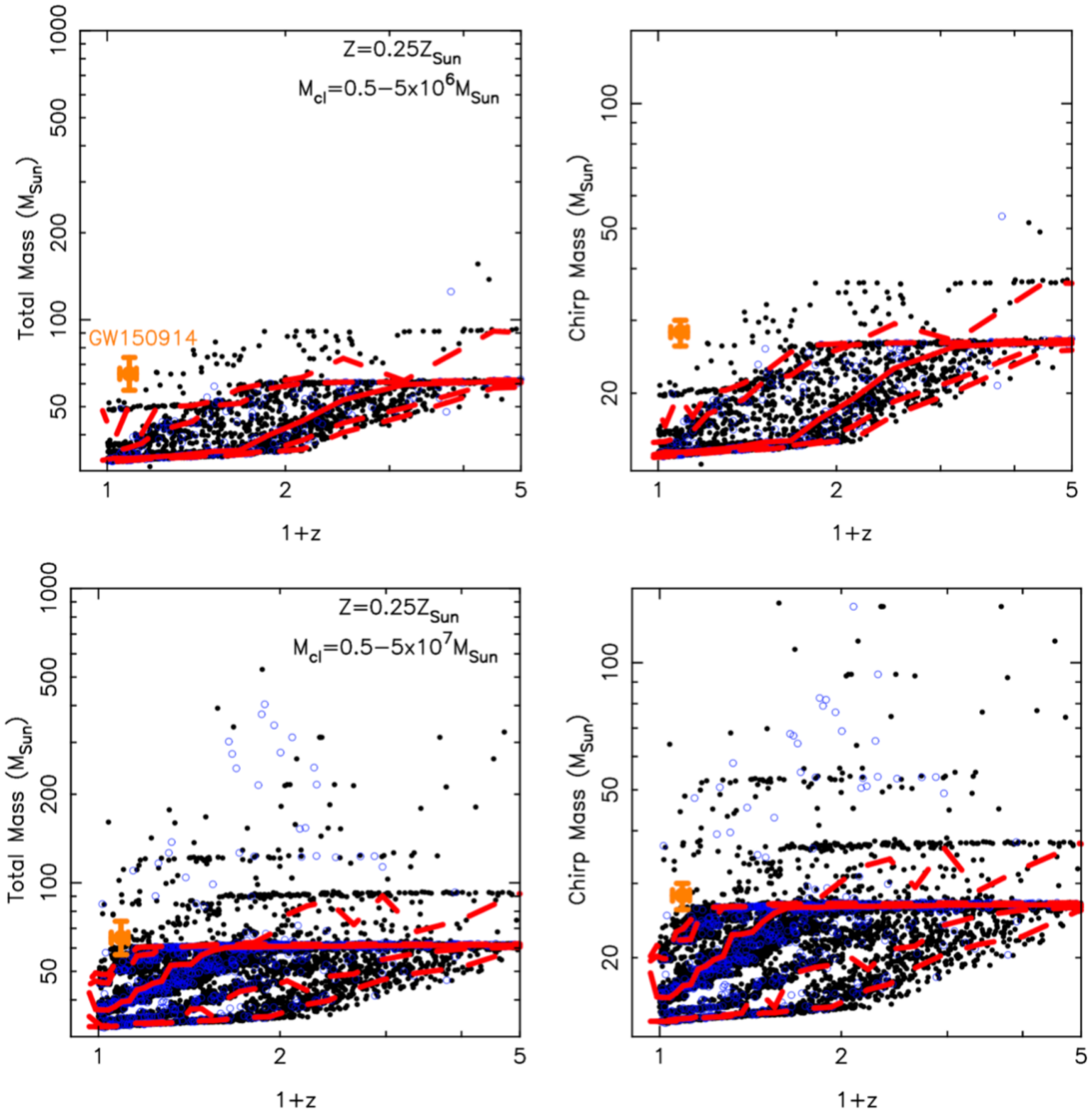}
\caption{Same as Figure\ \ref{Fig4} but for $Z=0.25\ Z_\odot$.
  }\label{Fig5} 
\end{figure*}

\subsection{Results}\label{results}
Given that our prescriptions are simplified in many ways, we
proceed here by testing the results of our models against the
results from the Monte Carlo models of \citet{2016arXiv160202444R}.

In the upper panel of Figure\ \ref{Fig3} we show the total number of
mergers per  cluster for systems  containing different numbers of stars  and
having different metallicities. In order to convert $M_{\rm cl}$ in
number of stars we have taken a mean stellar mass of $0.55\rm\
M_{\odot}$ typical of old stellar populations \citep{david-book}.
Moreover, we select here the BH spins based on the uniform
spin model described in Section \ref{gwr}.

Our semi-analytical models predict that the total number of mergers increases with
cluster mass and so do the Monte Carlo models.
The total number of inspirals over 12 Gyr is nearly linearly 
proportional to the final cluster mass. This
result is also in agreement with previous models of GCs and
shows that this statement can likely be extrapolated up to numbers of stars of 
order of a few $10^7M_{\odot}$.
Our models also predict an inversion of this simple correlation showing
that for $M_{\rm cl} \gtrsim 10^7M_{\odot}$ the number of merging BHs flattens or
even declines towards larger cluster masses. This is expected
given that for such massive clusters with higher values of 
$\sigma$ have a larger binary formation time $t_{1-2}$.
The most massive clusters in our integrations, which could represent NSCs, produce up to a few
thousand BH mergers per cluster. 

The lower panel of Figure\ \ref{Fig3} gives the 
median mass of the ejected BH binaries 
 formed in 10 cluster models with mass $1.2\times 10^6\ M_\odot$ and
half-mass radius $r_h=7\rm\ pc$.
These results are directly compared to those from
 Figure\ 4 in \citet{2016arXiv160202444R} which corresponds to a cluster model 
of  initial mass $1.2\times 10^6\ M_\odot$ and final half-mass radius
 $r_h\approx 7\rm\ pc$. The good agreement 
between the results of our simplified approach and those
of Monte Carlo simulations gives a high  level of
reliability to our semi-analytical models. \

 Figure\ \ref{Fig4} and Figure \ref{Fig5} give the masses for each of the BH inspirals
occurring  in  10  star cluster models of  metallicities
$Z=0.01 Z_\odot$ and  $Z=0.25\ Z_\odot$ respectively. Masses in the ranges
$0.5-5\times 10^6\ M_{\odot}$ (upper panel) and 
 $0.5-5\times 10^7M_{\odot}$ (lower panel) were considered. 
The overall structure of the plots agrees 
well with our understanding of the dynamics of BHs and their evolution
in star clusters,
and  with the results of previous work \citep[e.g., see Fig. 7 of][]{2016arXiv160202444R}.
After the formation of the cluster at high redshift, the BHs segregate
to the center,
the most massive BHs form binaries and 
the majority of them are ejected. The cluster processes through its 
 population of BHs that merge and are ejected 
from most to least massive, so that only low-mass BHs 
are retained by the present epoch.
More massive clusters,
which could represent typical NSCs,
 produce BH mergers in
 the local universe that are significantly more massive than mergers
 occurring in lower mass clusters. 

As also noted in \citet{2016arXiv160202444R} 
the plateaus in the chirp mass and total mass distributions  in 
Figure \ref{Fig5}
are mainly
a consequence of the maximum BH mass in the initial models, which is
regulated by the wind-driven mass loss from the Vink prescription.
For the high metallicity models  this 
produces a large population of $30\ M_\odot$ BHs, which 
leads to the formation of a large population of equal-mass mergers
 with total mass of $60\ M_\odot$. 
More interestingly, we find that massive cluster models produce an
additional collection of binaries at $90\ M_\odot$ 
and $120\ M_\odot$ 
which can be clearly seen
at high redshift in  Figure \ref{Fig5}. One of the
 two BHs in these binaries  has experienced 
one and two earlier mergers with lower mass BHs respectively.
 For the  the low metallicity models  there is no 
 apparent collection of sources as might be expected
 \citep{2016arXiv160202444R}. 
{ The decreased efficiency of the stellar winds in the 
low metallicity models implies that a lower 
number of high-mass stars are converted
 into BHs with the maximum-mass set by the  wind-driven mass loss
 prescription, resulting in a wider 
 range of BH masses.}

In  Figure\  \ref{Fig4} and Figure\  \ref{Fig5} we  show the  total and chirp mass and uncertainties 
associated with the recent detection of 
the  BH binary merger GW150914 \citep{2016PhRvL.116f1102A}. 
The reported masses of GW150914 are
consistent with the masses of black hole mergers from GCs in the local
universe. However, even for the low metallicity models 
only $5$ percent of the total  number of mergers
in GCs produce a merger at low redshift with
a total mass significantly larger than $50\ M_\odot$ as required to match the total
mass  of the GW150914 event.   In NSCs this percentage is
significantly  larger, being $\approx 20$ percent of the total number of
inspiraling binaries. In high metallicity clusters (Figure\  \ref{Fig5})  
a smaller number of high mass BH mergers is produced at low redshift making these clusters
less likely progenitors of GW150914-like events.

{ In Table 1 we report the mean number of mergers per cluster  obtained from our models.}
NSCs are defined here as clusters with masses in the range
$5\times 10^6 - 5\times 10^7 \ M_\odot$, while GCs have masses in the
range $10^5 -10^7 \ M_\odot$. 
In order to obtain the mean rate of  
mergers we
weighted the number of mergers from each of the cluster models
by  a cluster initial  mass function (CIMF). For GCs 
we assume a power law CIMF: ${\rm d} M/{\rm d} N\propto
M^{-2}$ \citep[e.g.,][]{2003A&A...397..473B}. For NSCs the initial mass function
is largely unknown. 
Here we take the IMF of NSCs directly from the 
mass distribution of NSCs at $z=2$ from the galaxy formation models
of \citet{2015ApJ...812...72A} (their Figure 10).  
These models produce a mass distribution at $z=0$ that is consistent with the
observed NSC
mass  distribution from \citet{2016MNRAS.457.2122G}.
 We note that here we might be underestimating the
number of {\it massive} mergers from NSCs occurring at low redshift because we have assumed
that  these systems are as old as Galactic GCs. 
In fact, while
most NSCs appear to be dominated by old stellar components 
they are also known to have a complex star formation history and to
 contain young stellar populations which can produce high mass
mergers also at later times (we will come back to this point below).
It is also possible that a large fraction of the NSC stars accumulated 
gradually in time
by infalling globular clusters that decayed to the center
through dynamical friction.
If this process is the main mechanism for NSC formation, then
NSCs and GCs will comprise similar stellar populations \citep{2014ApJ...794..106A}.

Table 1 shows that our models predict a few thousands BH mergers per
NSC over 12 Gyr of evolution.
This expectation also
appears to be  consistent with previous estimates
\citep{2000ApJ...528L..17P,2009ApJ...692..917M}.
In addition, NSCs produce between $50$ to $\approx 500$  BH mergers with high mass
$>50M_{\odot}$  at $z<0.3$ depending on the BH spin magnitudes and assumed
metallicities distribution of the
underlining stellar population.
Our GC models  produce only a few mergers per cluster within $z<0.3$ and
total mass $>50\ M_{\odot}$. These massive binaries are found to form 
only in the most massive GCs ($M_{\rm cl}\gtrsim
10^6M_{\odot}$) of  low metallicity.

The number of massive mergers at 
low redshift is also sensitive to the spin magnitude distribution we assume.
For high spin models,
 a smaller number of BHs are retained 
in the clusters compared to the uniform spin models.
Consequently, high spin models
produce fewer high mass BH mergers at low redshift compared to
models that assume low spins. However, in either spin models
a  number of inspiraling BH binaries with
mass $\gtrsim 50\ M_\odot$ is found to merge at low redshift.
Finally, Table 1 gives the number of BH mergers that are retained inside the cluster.
Between  10 and $20$ percent of high mass ($>50\ M_\odot$)  mergers occurring in NSCs at 
$z<1$ are retained inside the cluster enabling the formation
of even more massive BH mergers.

The results presented in this section suggest that NSCs are a natural
environment for producing BH mergers that are observable by aLIGO
detectors. In addition to this, NSCs can form high mass BH binaries, and 
mergers with mass consistent with that of GW150914 also in relatively high 
metallicity environments.
The implications of our results are discussed in more detail 
in the following section.

\begin{table*}
\caption{The mean number of inspirals per cluster over 12 Gyr of
  evolution, $\langle N \rangle$, that occur at redshift $z< 0.3$ and $z< 1$
and those that occur at redshift $z< 0.3$ and $z< 1$
and have a total mass $>50\ M_\odot$. Below we give the fraction of mergers
that are retained inside the clusters.}
\centering
\begin{tabular}{lllllll}
\hline
Model & $\langle N \rangle$ & ~
&  ~ \\
\hline
~ & total &$z<0.3$&  $z<1$ & $z<0.3,M>50\ M_\odot$ & $z<1,M>50\ M_\odot$ \\ 
~ & Z=0.01 (Z=0.25)  &  Z=0.01 (Z=0.25) & Z=0.01 (Z=0.25) & Z=0.01 (Z=0.25)&  Z=0.01 (Z=0.25)\\  
\hline
NSCs - low spins & 1379 (1307)  & 462 (450)& 990 (980) & 444 (166) & 877 (506)\\ 
GCs - low spins  & 96 (80) &  23 (20)& 57 (42) & 5 (0.5) & 20 (3) \\
NSCs - high spins& 1667  (1517) & 576 (510) & 1080 (1060)  & 423 (115)& 960 (490)  \\ 
GCs - high spins   & 109 (80) & 24 (20)& 55 (47) & 4 (0.2)& 20 (2)  \\
\hline
 IN-CLUSTER MERGERS \\ 
\hline
NSCs - low spins  & 0.2 (0.2)& 0.2 (0.2)& 0.2 (0.2)& 0.2 (0.2)& 0.2 (0.2)\\ 
GCs - low spins  &  0.07 (0.07) &  0.07 (0.07)& 0.07 (0.07) &  0.04 (0.03)& 0.07 (0.05) \\
NSCs - high spins &  0.07 (0.07) &  0.07 (0.07)& 0.07 (0.07) & 0.1 (0.1)&  0.1 (0.1)\\ 
GCs - high spins  &  0.02 (0.02) & 0.03 (0.03)& 0.03 (0.03)&  0.03 (0.04)& 0.03 (0.02)  \\
\hline
\end{tabular}
\label{t1}
\end{table*}

\section{Implications and Discussions}\label{disc}
Our study  shows that 
a multitude of BH binary mergers can 
 be produced at the center of galaxies where NSCs reside.
In the following we derive an approximate expected detection rate from
the results of our models and discuss some implications 
for  possible aLIGO detections of these mergers over the next decade.
Finally we discuss the production of BH mergers in NSCs hosting 
a central MBH, and the possibility of a continues supply of BH-binaries in NSCs
through episodic and/or continuous star formation.

\subsection{Detection rate estimates}
Here we use results from semi-analytical galaxy formation models to derive an 
expected MBH occupation fraction in NSCs and use this as 
well as the results of the cluster semi-analytical models presented in this paper 
to make predictions about the merger rate of BH binaries produced in NSCs.
We also consider the merger rates from GCs and compare these to 
estimates made in former studies.

{ To compute the aLIGO merger rate of BH binaries per unit volume  we 
use the following  expression:
\begin{equation}\label{r-f}
\Gamma^{\rm NSC}_{\rm aLIGO}=n_{\rm gx}\Gamma_{\rm merge}f_{\rm nucleated}
\end{equation}
where $n_{\rm gx}$ is 
the number density of galaxies, $\Gamma_{\rm merge}$ is the averaged merger rate of 
BH binaries per cluster that merge within the observable volume,
and $f_{\rm nucleated}$ is the fraction of galaxies which host a NSC but do not have a
MBH.}

\begin{figure} 
\centering 
\includegraphics[width=3.in,angle=270.]{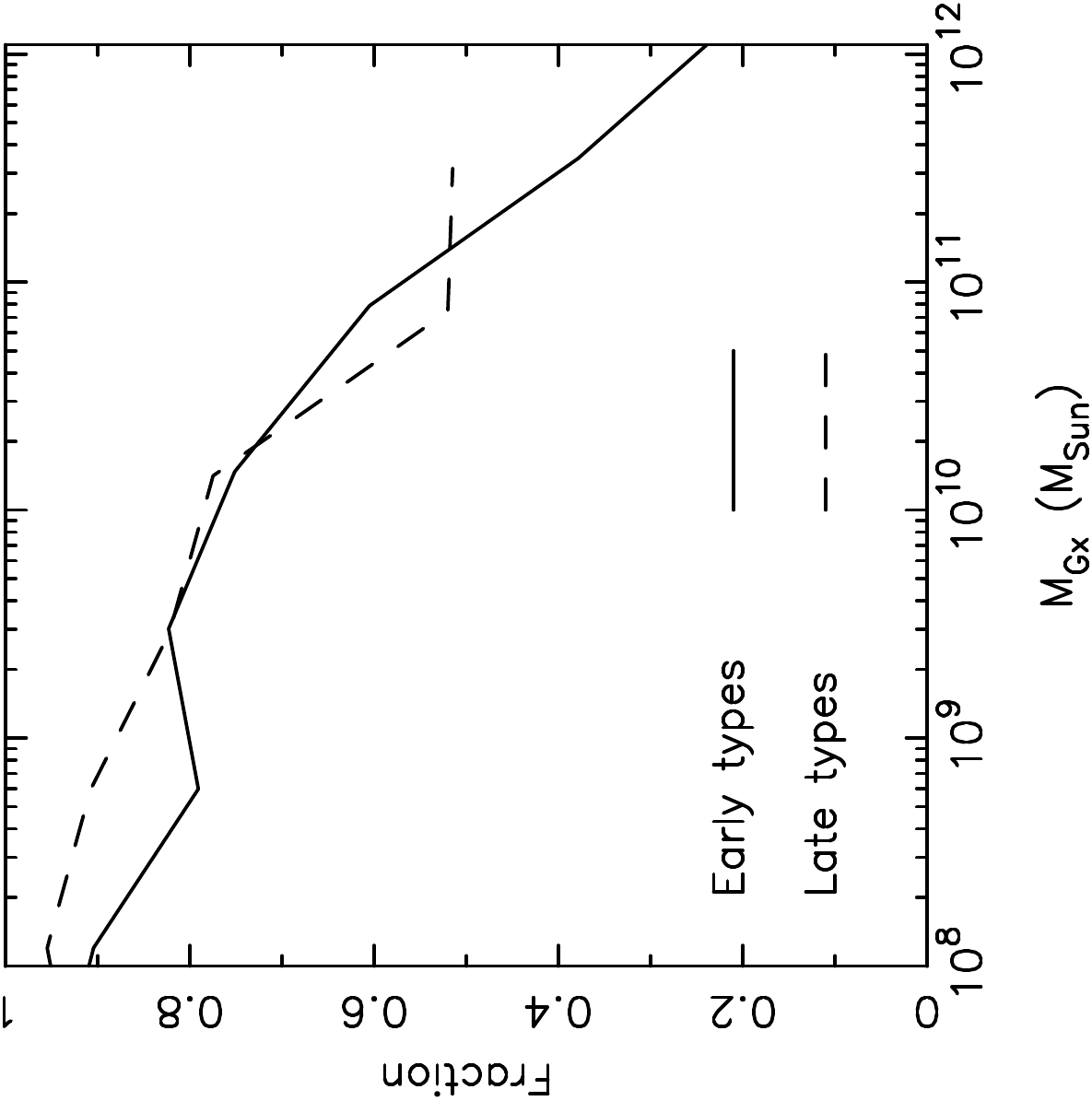}
\caption{Fraction of galaxies containing a NSC ($M_{\rm cl} \geq 10^5 \ M_\odot$)  but no MBH as a function of
\emph{total} galaxy mass.
Results are from the galaxy formation
models of \citet{2015ApJ...806L...8A} and \citet{2015ApJ...812...72A}.
Dashed line is for  galaxies with bulge-to-total-mass-ratio smaller than 0.7. Solid line is for early type galaxies which are  defined here as galaxies with  bulge-to-total-mass-ratio   larger than 0.7.
  }\label{Fig6} 
  \end{figure}

 While observations show that NSCs and MBHs coexist in some galaxies, and
that NSCs exist in most galaxies, $f_{\rm nucleated}$ remains largely 
unconstrained.
 Here we use the results of semi-analytical galaxy formation
models that follow the cosmological evolution of galaxies, their MBHs and NSCs. These 
galaxy formation models are described in   \citet{2012MNRAS.423.2533B}, \citet{2015ApJ...806L...8A} and \citet{2015ApJ...812...72A}.
Figure\ \ref{Fig6} shows the fraction of galaxies in these models that contain 
a NSC but do not host a MBH. These models predict that the number 
of galaxies hosting a NSC but without a MBH is quite large, being $f_{\rm nucleated}\gtrsim 0.5$
for galaxies with total mass $M_{\rm Gx}\lesssim 10^{11}\ M_\odot$ regardless
of galaxy type. Based on Figure \ref{Fig6} we adopt here a conservative 
value of $f_{\rm nucleated}= 0.5$, and {
adopt  a number density of  galaxies  of
$0.02\rm\ Mpc^{-3}$ \citep[e.g.,][]{2005ApJ...620..564C,2008ApJ...675.1459K}. }
Assuming that  BH-BH mergers can be seen by aLIGO 
out to a redshift of  $z\lesssim 0.3$ \citep[][]{2016ApJ...818L..22A}, 
which   corresponds to an age of the universe  of $\approx 10 \rm\ Gyr$ \citep{2015arXiv150201582P},
then Eq.\ (\ref{r-f})
gives a merger rate of BHs in NSCs of
\begin{equation}\label{gnsc}
\Gamma_{\rm aLIGO}^{\rm NSC}\approx 1.5 \rm\ Gpc^{-3}\  yr^{-1}.
\end{equation}
Thus our calculation predicts a substantial number of detectable BH mergers from 
NSCs. From Table 1 we see also that between
10 and  $20$  percent of the total number of merger remnants 
are retained inside NSCs.
Adopting an aLIGO
detection-weighted comoving volume of $\approx \rm 10{\rm\ Gpc^{-3}}$ for full design sensitivity
\citep{2016ApJ...818L..22A}, the rate in Eq. (\ref{gnsc})
 translates into a detection rate of $\approx 10 \rm\  yr^{-1}$.

The merger rate of Eq.\ (\ref{gnsc}) can be directly compared to that from GCs:
\begin{equation}\label{gcs}
\Gamma_{\rm aLIGO}^{\rm GC}\approx 5  \rm\ Gpc^{-3}\  yr^{-1},
\end{equation}
which was obtained from the number of mergers per GCs within $z<0.3$ in Table 1 and assuming a
number density  of GCs equal to $0.77\rm Gpc^{-3}$ \citep{2015PhRvL.115e1101R}.
Note that Eq.\ (\ref{gcs}) is very well consistent with the rate previously 
derived by other authors \citep{2016arXiv160202444R}.
As before, taking an aLIGO
detection-weighted comoving volume of $\approx 10\rm\ Gpc^{-3}$ for full design sensitivity,
we obtain an aLIGO detection rate of BH mergers from GCs of $\approx 50\rm\  yr^{-1}$.

Now we only consider mergers with mass $\gtrsim 50M_{\odot}$ 
occurring at a redshift  $z<0.3$ (see Table 1). We define these as 
possible progenitors of  the event GW150914. 
The detection rate
of high mass mergers from NSCs in the local universe  is in the range
\begin{equation}\label{ncs2}
\Gamma_{\rm aLIGO}^{\rm NSC}(z<0.3;M>50M_{\odot})
\approx 0.4 -1  \rm\ Gpc^{-3}\  yr^{-1},
\end{equation}
where the lower limit corresponds to high metallicity clusters and the high spin model, 
and the upper limit to low metallicity clusters and to a uniform BH spin distribution.
Interestingly, we find that
between 20 and 50 percent of all massive mergers in NSCs are produced by the consecutive merger
channel discussed in this paper with a few percent having a mass $\gtrsim 100M_{\odot}$.
The corresponding merger  rate 
of high mass BH binaries mergers in GCs  is
\begin{equation} 
\Gamma_{\rm aLIGO}^{\rm GC}(z<0.3;M>50M_{\odot})\approx 0.05 -1
\rm\ Gpc^{-3}\  yr^{-1} \ ,
\end{equation}
similar to that corresponding to high  mass BH mergers 
produced in  NSCs. Both rates of high mass mergers from GCs and NSCs
are  marginally consistent with the rate of $2-53\rm\ Gpc^{-3}\  yr^{-1}$
of GW150914-like mergers given  by \citet{2016arXiv160203842A}.

From our rate computation we find that the detection rate of BH-BH binaries 
from NSCs is substantial, and it is about one tenth  of that from GCs.
Importantly, we also find that the NSC detection rate of high mass BH mergers similar to GW150914 
($M\approx 60\ M_\odot,z<0.5$)  is comparable to that from GCs, with many of the mergers
being produced through the consecutive merger scenario discussed in this paper.
Finally, our results show that most   GW150914-like mergers are
more likely to be a product of dynamical interactions occurring in massive clusters 
of low metallicity, also in agreement with previous findings \citep{2016arXiv160404254R}.

\begin{figure} 
\centering
\includegraphics[width=3.in,angle=270.]{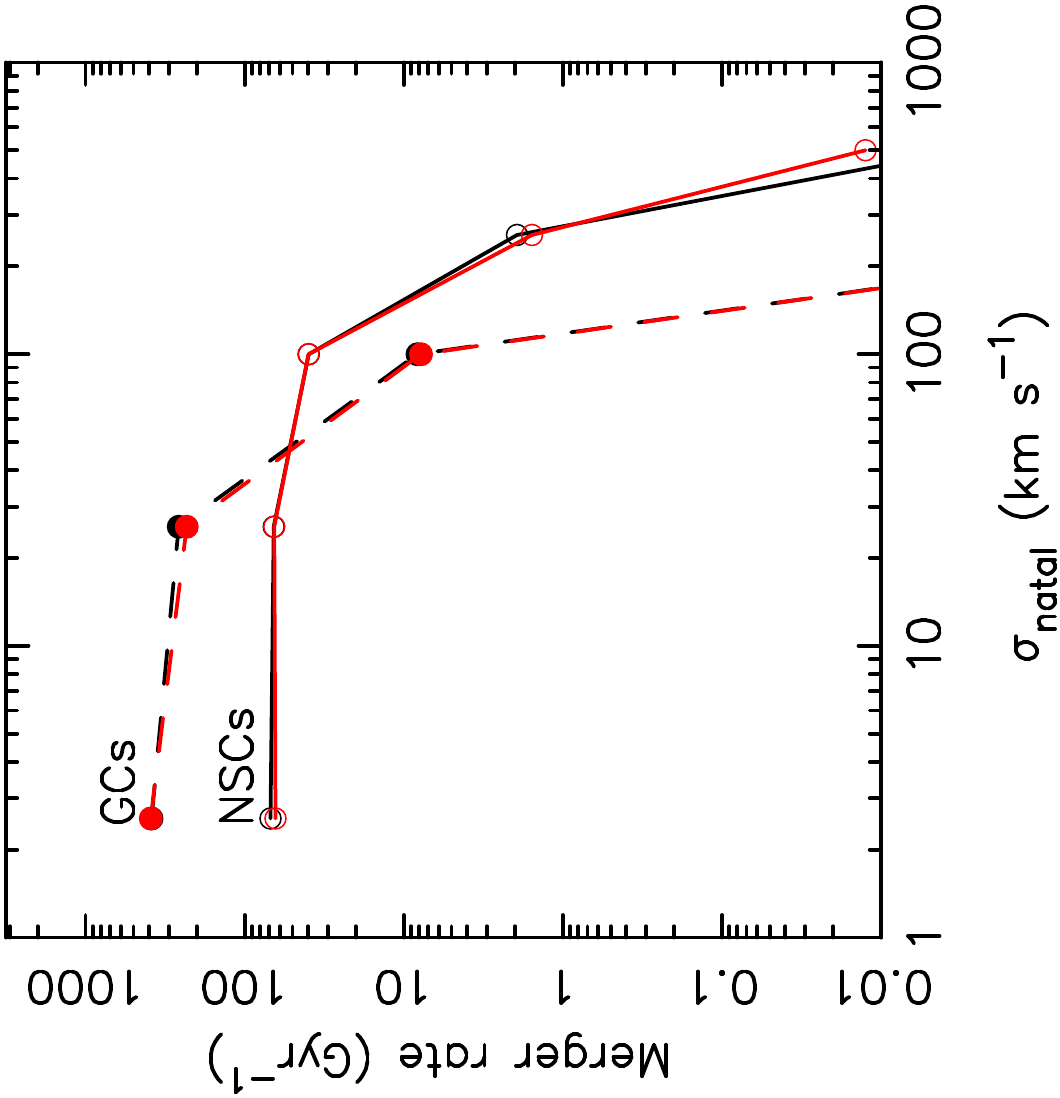}
\caption{Mean merger rate per galaxy at redshift $z<1$
as a function of  the dispersion 
of the Maxwellian distribution of natal kick velocities, $\sigma_{\rm natal}$,
applied to all BHs independently of mass.
Red curves  correspond to a
metallicity $Z=0.25\ Z_{\rm \odot}$  and 
black curves to a metallicity
$Z=0.01\ Z_{\rm \odot}$. 
We have assumed a number of GCs per galaxy equal to 100 
and a light travel time of $8\rm Gyr$ at $z=1$.
At $\sigma_{\rm natal}\gtrsim 50 \rm\ km\ s^{-1}$
the merger rate of BH binaries from NSCs is dominant compared to 
the corresponding merger rate from GCs.
  }\label{Fig5b} 
\end{figure}

\subsection{Dependence on natal kicks}\label{bkick}
\citet{2016ApJ...818L..22A} noted that 
``\emph{for both  dynamical formation in 
[globular]
 clusters and isolated binary evolution, the implication of BH binary existence is that BH natal kicks cannot always 
 be high ($ \gtrsim 100 \rm\ km\ s^{-1}$), in order to avoid disrupting or widening the orbits too much, or ejecting the BHs from clusters before they can interact.}''
For example, large natal kicks will widen the orbits of massive binary progenitors and 
quench  the formation of binary BH  systems that will 
merge within the age of the universe \citep[e.g.,][]{2007ApJ...662..504B}.
 \citet{2016ApJ...818L..22A} did not consider the possibility that BH binaries can
 be dynamically assembled inside NSCs which we discuss here.

Motivated by the fact that
 the distribution of formation kicks for BHs is largely uncertain, 
even at the qualitative level, we assume here that 
 the kick magnitude distributions are fully unconstrained and
explore how the NSC and GC BH merger rates 
are affected by varying 
these distributions.
In Figure \ref{Fig5b} we 
plot the mean merger rate per galaxy at $z<1$ in models 
where the  BHs are given a natal kicks taken
from Maxwellian distributions with dispersion 
$\sigma_{\rm natal}$, and the spin magnitude of the BHs
is uniformly  distributed in the range $\chi= [0, 1)$.
 Unlike 
the models discussed above, here we assume that the
BH natal kick distributions do not depend 
on the mass of the BHs. 

As $\sigma_{\rm natal}$ increases above $\gtrsim 20\rm\ km\ s^{-1}$
the merger rate from GCs decreases and becomes comparable to that
of NSCs for $\sigma_{\rm natal}\approx 50\rm\ km\ s^{-1}$.
Above this  value  the merger rate of NSCs
begins to decrease as well because a fraction of BHs 
starts to be ejected from  low mass NSCs as well. However, for $\sigma_{\rm natal}\gtrsim 50\rm\ km\ s^{-1}$
we 
see that the BH merger rate from NSCs becomes significantly larger than the 
corresponding merger rate from GCs.
We conclude that for high natal kicks, $\gtrsim 50\rm\ km\ s^{-1}$,
NSCs can dominate the merger rate of 
dynamically formed BH binaries that are detectable by aLIGO. 

 Recent theoretical studies  suggest that the birth kicks may not be directly 
 correlated with the BH mass \citep{2015MNRAS.453.3341R,2015ApJ...801...90P}, and 
that  the distribution of black hole kick velocities 
 could be similar to that of 
 neutron stars $\sigma_{\rm natal} \approx 200 \rm\ km\ s^{-1}$ \citep[but see][]{2016MNRAS.456..578M}. 
 If this is the case, then the merger rate of BH binaries from NSCs
will greatly  dominate over that from GCs.
However, 
we also note that 
 for  $\sigma_{\rm natal}\gtrsim 200\rm\ km\ s^{-1}$
the rate of BH mergers from NSCs becomes significantly smaller
than the rate of $2-53\rm\ Gpc^{-3}\  yr^{-1}$
implied by the detection of GW150914 \citep{2016arXiv160203842A}.
Thus very high values of natal kicks are 
excluded by our analysis when combined with the estimated aLIGO detection rate.

\subsection{Mergers of stellar remnants near massive black holes}
So far we have considered the merger of BH binaries in stellar clusters
which do not host a central MBH. This is certainly not the case for   at least 
a handful of NSCs which are known to have MBHs at their center \citep[e.g.,][]{2008ApJ...678..116S}.
One example is our own Galaxy
which contains a $\approx 10^7M_{\odot}$ NSC and  a central 
MBH of $\approx4\times  10^6\ M_\odot$ \citep{2008ApJ...689.1044G,2009ApJ...692.1075G}.
More generally NSCs and MBHs are known to co-exist in galaxies with masses
$\approx 10^{10}\ M_\odot$ \citep[e.g.,][]{2008ApJ...678..116S,2008AJ....135..747G};
galaxies with masses lower than this value show clear evidence for nucleation but little evidence for a MBH. Conversely galaxies with masses above $10^{11}\ M_\odot$
 are dominated by MBHs but generally show no evidence for nucleation \citep{2012AdAst2012E..15N}.

In NSCs containing a central MBH, the merger rates of BH binaries
are expected to be significantly different, although not necessarily smaller, than the rates
given in Table 1. If a MBH is present it will inhibit core-collapse, causing the formation of a 
 Bahcall-Wolf cusp instead \citep{1976ApJ...209..214B}. After the BHs
 have segregated to the center their densities will dominate over the stellar densities 
within a radius $\lesssim 0.1$ times the MBH influence radius
\citep{2006ApJ...645L.133H,2012ApJ...744...74G} --
 note that this latter statement  depends on the formation history of the NSC \citep{2014ApJ...794..106A}.
At such small distances from the MBH all binaries will be effectively soft so that any interaction 
with a third BH or star will tend to make the binary internal orbit wider.
In this situation, three-body interactions between binaries and field objects will not lead to BH mergers but rather
to the  ``evaporation'' of binaries.

Several mechanisms have been discussed in the literature which can produce  BH mergers even in the extreme stellar environment of
MBHs. Below
we briefly review two of these processes: (i) mergers of BH binaries due to Lidov-Kozai (LK) resonance induced by the central MBH \citep{2012ApJ...757...27A}; 
(ii) single-single captures of compact objects due to 
gravitational wave energy loss \citep{1990ApJ...356..483Q,1993ApJ...418..147L,2009MNRAS.395.2127O}.

\subsubsection{MBH-mediated BH mergers in NSCs}
\citet{2012ApJ...757...27A} showed that near a MBH
the dynamical  evolution of binaries is dominated by perturbations from
the central MBH. In particular, the LK
mechanism \citep{1962P&SS....9..719L,1962AJ.....67..591K} exchanges the relative inclination
of the inner  BH binary orbit to its orbit around the MBH for the eccentricity of
the BH binary \citep{2010ApJ...713...90A}. At the peak eccentricity, during a Lidov-Kozai cycle, GW emission can become efficient leading to a merger of the two BHs.

More recently \citet{2016arXiv160404948V} performed  $N$-body simulations
of small clusters of stars containing  a central MBH, to estimate the rate of BH mergers
induced by the LK mechanism. These authors found that 
this mechanism could produce mergers at a maximum 
rate of $\approx 2$ per Myr per Milky Way equivalent galaxy. This rate
appears to be
somewhat comparable to the upper end of the expected detection rate from 
stellar clusters \citep{2016arXiv160202444R}.
\citet{2016arXiv160404948V}  showed that this rate could translates into
a maximum rate per volume of $\approx 100\rm\ Gpc^{-3}\  yr^{-1}$.
However, as also noted by \citet{2016arXiv160404948V}  
their merger rate estimates are likely to be an 
overestimate of the true  merger rate from the LK process as they used optimistic values for both the merger fraction as well as for the BH binary fraction.

The efficiency of  the LK process in inducing BH mergers in  NSCs with MBHs is sensitive to major uncertainties. 
The major challenge to this being that a continuous supply
of BH binaries is needed in order to obtain a finite merger
rate.
In fact, a  continuous formation of binaries is part of the assumptions made in the rate estimates of \citet{2012ApJ...757...27A} and \citet{2016arXiv160404948V} .

Binaries well inside the influence radius of a MBH will 
be essentially all soft and will
be disrupted  over the  typical timescale \citep{BT:87}:
\begin{eqnarray}
t_{\rm ev}&=&{m_{12}\sigma \over
16\sqrt{\pi} G m_\star \rho a \ln \Lambda} 
\approx 10^7 
{\sigma\over100 \rm\ km\ s^{-1}}
 \\
&& 
\left(0.5 \frac{m_{12}}{m_\bullet}\right)
\left(\frac{\ln \Lambda}{10} \frac{\rho}{10^6 \ M_\odot {\rm\ pc^{-3}}} 
{a\over\rm 1\ AU} \right)^{-1}.~~~~\nonumber 
\end{eqnarray}
Because $t_{\rm ev}$ is much shorter than the  lifetime of any NSC,
we expect that most primordial binaries will be disrupted by now
in these systems.
  \citet{2012ApJ...757...27A} describe various processes 
which may affect the replenishment rate of compact binaries and/or their progenitors in NSCs with a MBH. \citet{2009ApJ...698.1330P} suggested that disruption of triple stars could leave behind a binary in a close orbit around the MBH and could serve as a continuous source of replenished binaries close to the MBH. 
In addition, in-situ star formation can also repopulate the binary population of NSCs.
Central star formation bursts may occur continuously (or episodically) throughout the evolution of the stellar cusp 
and lead to
a steady population of massive binaries near the center. 
Finally, NSCs might result from the merger of stellar clusters
in the inner galactic regions \citep[e.g.,][]{2015ApJ...812...72A}.  
Such clusters may harbor an inner core cluster of BHs that formed during the cluster evolution. If these BHs are retained in the cluster this mechanism may also
 contribute to the BH and BH-binary populations in NSCs \citep{2014ApJ...794..106A}.

\subsubsection{Mergers from BH-BH scattering in NSCs}
\citet{2009MNRAS.395.2127O} showed that
in the dense stellar environments 
such as those of NSCs BH binaries
can efficiently form out of GW radiation during
BH-BH (single-single) encounters. Interestingly 
they show that most of the mergers from this channel 
will have a finite eccentricity while they enter
the $10\rm\ Hz$ frequency band of aLIGO.
This processes could become important for sufficiently large cluster masses.
However, the predicted rate of
 BH-BH mergers from this channel, although very uncertain, 
is estimated to be only  
$\sim 0.01\rm\ Gpc^{-3}\  yr^{-1}$  \citep{2013ApJ...777..103T}, and
therefore it is sub-dominant with respect
to the rate from the other processes discussed in
this paper. Finally, we note that
\citet{2016ApJ...816...65A} argued 
that even the rate of \emph{eccentric} mergers from 
BH-BH scattering in NSCs is likely to be much smaller
to that of eccentric mergers from BH triples
formed in GCs.

In conclusion,  the role of NSCs containing MBHs in producing BH mergers is 
a subject of considerable interest, which will likely require high precision $N$-body simulations
of large number of particles. For now, theoretical models suggest that the merger rate of
BHs in these systems might be considerably lower than that from NSCs without MBHs.

\subsection{Continuous and episodic star formation in NSCs}
As mentioned above, most NSCs are known to have undergone
a complex star formation history characterized by recurrent episodes 
of star formation \citep[e.g.,][]{2006ApJ...649..692W}.
Thus NSCs  might still be forming BHs  and BH binaries at the present epoch.
This is different from GCs where all stars are old (ages $\gtrsim 10^{10}\rm \ yr$)
and many (if not all) BHs are expected to have been already  ejected  by now.

As an example,
our Galactic center contains a large population of young 
massive stars, many of which reside in a stellar disk. These stars most likely 
originated in-situ following the fragmentation of a gaseous disk formed from
 an infalling gaseous clump \citep[e.g.,][]{2008Sci...321.1060B}. Such stars formation bursts 
are thought to occur episodically throughout the evolution of the central cusp. Eclipsing and close binaries are observed among the Galactic center young stars, suggesting star formation as an additional process which can repopulate the binary population (including BH binaries) in the 
NSC \citep{2014ApJ...782..101P}.

Observational studies of NSCs in external galaxies, including high resolution spectroscopic surveys,
have been used to characterize the star formation
 history and ages of NSCs \citep{2006ApJ...649..692W,2006AJ....132.1074R}.
 The common finding emerging from these studies is that most NSCs 
are 
 characterized by a mixture of morphological components and different stellar populations spanning a wide range of characteristic ages from 10 Myr to 10 Gyr. 
 Observations also suggest that the ages of NSCs and masses  depend on the host galaxy Hubble type, with NSCs in early-type spirals being older and more massive than those of late-type 
 spirals \citep{2006AJ....132.1074R}. 
 More generally the growth of the nuclei might be  a continuous and ongoing process occurring during and after most of the host galaxy was formed. 
 
 As shown in Figure \ref{Fig4} and \ref{Fig5},
 most massive BH mergers ($\gtrsim 50 M_{\odot}$) 
occur at early times in the lifetime of a single stellar population star cluster.
This is because 
the most massive objects segregate earlier and
are ejected earlier through dynamical interactions. This will not be the case 
if the clusters form new stellar populations at later times. New episodes of star formations
in NSCs will lead to the formation of new BHs with a substantial contribution to 
the merger rate  in the local universe. 
 In-situ star formation could therefor  contribute significantly to the detection rate of 
high mass mergers we previously derived (c.f., Eq. [\ref{ncs2}]),
although it will likely not affect the total detection  rate of BH mergers in NSCs.

\section{Summary}\label{summ}
Understanding the distribution of BHs at the centers of galaxies  is crucial for making predictions about the expected event rate and  source properties
for high-frequency gravitational wave detectors.
 Since the distribution of stellar BHs is not known, 
and Monte Carlo simulations of  star clusters  are still 
limited to a few $10^6$ particles, we opted here for a
 semi-analytical approach which we used in order to make predictions about the
 properties and rate of BH binary mergers that are dynamically assembled in NSCs.
 In the future we plan to explore this topic  using more accurate,
although  computationally  more demanding,  Monte Carlo simulations.
The main conclusions of our work are summarized below.

\begin{itemize}
\item[1)] NSCs produce BH binary mergers at a realistic rate of 
$\approx 1.5\rm\ Gpc^{-3}\  yr^{-1}$. 
\item[2)] BHs in NSCs can experience a number of mergers and 
grow to masses up to a few hundred solar masses. 
Although rare, such  high-mass BH mergers at low redshift are unique to NSCs, because these
are the only clusters with sufficiently high escape velocities such that they can retain a large 
fraction of their merging BHs.
\item[3)] Assuming that BHs receive low natal kicks, with
an imparted  momentum equal to the momentum imparted to neutron stars,
then the NSC detection rate of high mass BH mergers similar to GW150914
 ($M \geq 50\ M_\odot ,\ z \leq 0.3$) is $0.4-1\rm\ Gpc^{-3}\  yr^{-1}$. This rate
 is  comparable or larger tan the
 corresponding merger rate  of dynamically formed BH binaries in GCs.
\item[4)] If BHs receive natal kicks as large as
$\gtrsim 50\rm\ km\ s^{-1}$
then BH binary mergers produced dynamically in NSCs
could dominate over the merger rate of similar sources produced either in  GCs
or through isolated binary evolution.
\end{itemize}

\bigskip
We thank Enrico Barausse, Sourav Chatterjee, Cole Miller, Carl Rodriguez
 for useful discussions and the anonymous referee for useful suggestions. 
FA acknowledges support from a CIERA postdoctoral fellowship at Northwestern University.
FAR acknowledges support from NSF Grant AST-1312945 and NASA Grant NNX14AP92G,
at Northwestern University, and from NSF Grant PHY-1066293 through the Aspen Center for Physics..


\begin{thebibliography}{67}

\bibitem[Arzoumanian et al.(2002)]{2002ApJ...568..289A} Arzoumanian, Z., Chernoff, D.~F., \& Cordes, J.~M.\ 2002, \apj, 568, 289 

\bibitem[Aarseth(2012)]{2012MNRAS.422..841A} Aarseth, S.~J.\ 2012, \mnras, 422, 841 

\bibitem[Abbott et al. (2016a)]{2016PhRvL.116f1102A} Abbott, B.~P., Abbott, R., Abbott, T.~D., et al.\ 2016, Physical Review Letters, 116, 061102  

\bibitem[Abbott et al. (2016b)]{2016arXiv160203840T} Abbott, B.~P., Abbott, R., Abbott, T.~D., et al. 2016, arXiv:1602.03840  

\bibitem[Abbott et al.(2016c)]{2016arXiv160203842A} Abbott, B.~P., Abbott, R., Abbott, T.~D., et al.\ 2016, arXiv:1602.03842  

\bibitem[Abbott et al.(2016d)]{2016ApJ...818L..22A} Abbott, B.~P., Abbott, R., Abbott, T.~D., et al.\ 2016, \apjl, 818, L22 

\bibitem[Antonini et al.(2010)]{2010ApJ...713...90A} Antonini, F., Faber, J., Gualandris, A., \& Merritt, D.\ 2010, \apj, 713, 90 

\bibitem[Antonini \& Perets(2012)]{2012ApJ...757...27A} Antonini, F., \& Perets, H.~B.\ 2012, \apj, 757, 27 

\bibitem[Antonini(2014)]{2014ApJ...794..106A} Antonini, F.\ 2014, \apj, 794, 106 

\bibitem[Antonini et al.(2015a)]{2015ApJ...806L...8A} Antonini, F., Barausse, E., \& Silk, J.\ 2015a, \apjl, 806, L8 

\bibitem[Antonini et al.(2015b)]{2015ApJ...812...72A} Antonini, F., Barausse, E., \& Silk, J.\ 2015b, \apj, 812, 72 

\bibitem[Antonini et al.(2016)]{2016ApJ...816...65A} Antonini, F., Chatterjee, S., Rodriguez, C.~L., et al.\ 2016, \apj, 816, 65 

\bibitem[Bahcall \& Wolf(1976)]{1976ApJ...209..214B} Bahcall, J.~N., \& Wolf, R.~A.\ 1976, \apj, 209, 214 

\bibitem[Banerjee et al.(2010)]{2010MNRAS.402..371B} Banerjee, S., Baumgardt, H., \& Kroupa, P.\ 2010, \mnras, 402, 371 

\bibitem[Barausse(2012)]{2012MNRAS.423.2533B} Barausse, E.\ 2012, \mnras, 423, 2533 

\bibitem[Belczynski et al.(2002)]{2002ApJ...572..407B} Belczynski, K., Kalogera, V., \& Bulik, T.\ 2002, \apj, 572, 407 


\bibitem[Belczynski et al.(2007)]{2007ApJ...662..504B} Belczynski, K., Taam, R.~E., Kalogera, V., Rasio, F.~A., \& Bulik, T.\ 2007, \apj, 662, 504 

\bibitem[Belczynski et al.(2010)]{2010ApJ...715L.138B} Belczynski, K., Dominik, M., Bulik, T., et al.\ 2010, \apjl, 715, L138 


\bibitem[Belczynski et al.(2016)]{2016ApJ...819..108B} Belczynski, K., Repetto, S., Holz, D.~E., et al.\ 2016, \apj, 819, 108 

\bibitem[Berti et al.(2007)]{2007PhRvD..76f4034B} Berti, E., Cardoso, V., Gonzalez, J.~A., et al.\ 2007, \prd, 76, 064034 

\bibitem[Bik et al.(2003)]{2003A&A...397..473B} Bik, A., Lamers, H.~J.~G.~L.~M., Bastian, N., Panagia, N., \& Romaniello, M.\ 2003, \aap, 397, 473 

\bibitem[Binney \& Tremaine~(1987)]{BT:87}
Binney J., \&Tremaine S., 1987, Galactic Dynamics. Princeton Univ. Press, New Jersey, USA

\bibitem[Blandford \& Hughes (2003)]{BH:03} Blandford and
Hughes, S. 2003, ApJ, 585, L101

\bibitem[Bogdanovi{\'c} et al.(2007)]{2007ApJ...661L.147B} Bogdanovi{\'c}, T., Reynolds, C.~S., \& Miller, M.~C.\ 2007, \apjl, 661, L147 

\bibitem[B{\"o}ker et al.(2004)]{2004AJ....127..105B} B{\"o}ker, T., Sarzi, M., McLaughlin, D.~E., et al.\ 2004, \aj, 127, 105 

\bibitem[Bonnell \& Rice(2008)]{2008Sci...321.1060B} Bonnell, I.~A., \& Rice, W.~K.~M.\ 2008, Science, 321, 1060 

\bibitem[Chandrasekhar(1943)]{1943ApJ....97..255C} Chandrasekhar, S.\ 1943, \apj, 97, 255 

\bibitem[Chatterjee et al.(2010)]{2010ApJ...719..915C} Chatterjee, S., Fregeau, J.~M., Umbreit, S., \& Rasio, F.~A.\ 2010, \apj, 719, 915 

\bibitem[Chatterjee et al.(2016)]{2016arXiv160300884C} Chatterjee, S., Rodriguez, C.~L., \& Rasio, F.~A.\ 2016, arXiv:1603.00884 

 \bibitem[Conselice et al.(2005)]{2005ApJ...620..564C} Conselice, C.~J., Blackburne, J.~A., \& Papovich, C.\ 2005, \apj, 620, 564 
 
\bibitem[C{\^o}t{\'e} et al.(2006)]{2006ApJS..165...57C} C{\^o}t{\'e}, P., Piatek, S., Ferrarese, L., et al.\ 2006, \apjs, 165, 57 

\bibitem[de Mink \& Mandel(2016)]{2016arXiv160302291D} de Mink, S.~E., \& Mandel, I.\ 2016, arXiv:1603.02291 

\bibitem[Do et al.(2015)]{2015ApJ...809..143D} Do, T., Kerzendorf, W., Winsor, N., et al.\ 2015, \apj, 809, 143 

\bibitem[Dominik et al.(2012)]{2012ApJ...759...52D} Dominik, M., Belczynski, K., Fryer, C., et al.\ 2012, \apj, 759, 52 

\bibitem[Dominik et al.(2013)]{2013ApJ...779...72D} Dominik, M., Belczynski, K., Fryer, C., et al.\ 2013, \apj, 779, 72 


\bibitem[Downing et al.(2011)]{2011MNRAS.416..133D} Downing, J.~M.~B., Benacquista, M.~J., Giersz, M., \& Spurzem, R.\ 2011, \mnras, 416, 133 

\bibitem[Duquennoy \& Mayor(1991)]{1991AA...248..485D} Duquennoy, A., \& Mayor, M.\ 1991, \aap, 248, 485 


\bibitem[Figer et al.(2004)]{2004ApJ...601..319F} Figer, D.~F., Rich, R.~M., Kim, S.~S., Morris, M., \& Serabyn, E.\ 2004, \apj, 601, 319 

\bibitem[Fryer \& Kalogera(2001)]{2001ApJ...554..548F} Fryer, C.~L., \& Kalogera, V.\ 2001, \apj, 554, 548 

\bibitem[Georgiev \& B\"{o}ker(2014)]{2014MNRAS.441.3570G} Georgiev, I.~Y., B\"{o}ker, T.\ 2014, \mnras, 441, 3570 

\bibitem[Georgiev et al.(2009)]{2009MNRAS.396.1075G} Georgiev, I.~Y., Hilker, M., Puzia, T.~H., Goudfrooij, P., \& Baumgardt, H.\ 2009, \mnras, 396, 1075 

\bibitem[Georgiev et al.(2016)]{2016MNRAS.457.2122G} Georgiev, I.~Y., B{\"o}ker, T., Leigh, N., L{\"u}tzgendorf, N., \& Neumayer, N.\ 2016, \mnras, 457, 2122 

\bibitem[Gebhardt et al.(2001)]{2001AJ....122.2469G} Gebhardt, K., Lauer, T.~R., Kormendy, J., et al.\ 2001, \aj, 122, 2469 

\bibitem[Ghez et al.(2008)]{2008ApJ...689.1044G} Ghez, A.~M., Salim, S., Weinberg, N.~N., et al.\ 2008, \apj, 689, 1044-1062 

\bibitem[Gillessen et al.(2009)]{2009ApJ...692.1075G} Gillessen, S., Eisenhauer, F., Trippe, S., et al.\ 2009, \apj, 692, 1075 


\bibitem[{{Gonz{\'a}lez} {et~al.}(2007){Gonz{\'a}lez}, {Sperhake},
  {Br{\"u}gmann}, {Hannam}, \& {Husa}}]{g07}
{Gonz{\'a}lez} J.~A., {Sperhake} U., {Br{\"u}gmann} B., {Hannam} M., {Husa} S.,
  2007, Physical Review Letters, 98, 091101

\bibitem[Gonz{\'a}lez Delgado et al.(2008)]{2008AJ....135..747G} Gonz{\'a}lez Delgado, R.~M., P{\'e}rez, E., Cid Fernandes, R., \& Schmitt, H.\ 2008, \aj, 135, 747 

\bibitem[Gualandris \& Merritt(2012)]{2012ApJ...744...74G} Gualandris, A., \& Merritt, D.\ 2012, \apj, 744, 74 

\bibitem[G{\"u}ltekin et al.(2004)]{2004ApJ...616..221G} G{\"u}ltekin, K., Miller, M.~C., \& Hamilton, D.~P.\ 2004, \apj, 616, 221 


\bibitem[G{\"u}ltekin et al.(2006)]{2006ApJ...640..156G} G{\"u}ltekin, K., Miller, M.~C., \& Hamilton, D.~P.\ 2006, \apj, 640, 156 

\bibitem[Hansen \& Phinney(1997)]{1997MNRAS.291..569H} Hansen, B.~M.~S., \& Phinney, E.~S.\ 1997, \mnras, 291, 569 

\bibitem[Harris(1996)]{1996AJ....112.1487H} Harris, W.~E.\ 1996, \aj, 112, 1487 

\bibitem[Heggie(1975)]{1975MNRAS.173..729H} Heggie, D.~C.\ 1975, \mnras, 173, 729 


\bibitem[Hobbs et al.(2005)]{2005MNRAS.360..974H} Hobbs, G., Lorimer, D.~R., Lyne, A.~G., \& Kramer, M.\ 2005, \mnras, 360, 974 

\bibitem[Hofmann et al.(2016)]{2016arXiv160501938H} Hofmann, F., Barausse, E., \& Rezzolla, L.\ 2016, arXiv:1605.01938 

\bibitem[Hopman \& Alexander(2006)]{2006ApJ...645L.133H} Hopman, C., \& Alexander, T.\ 2006, \apjl, 645, L133 


\bibitem[Hurley et al.(2002)]{2002MNRAS.329..897H} Hurley, J.~R., Tout, C.~A., \& Pols, O.~R.\ 2002, \mnras, 329, 897 

\bibitem[Hurley et al.(2007)]{2007ApJ...665..707H} Hurley, J.~R., Aarseth, S.~J., \& Shara, M.~M.\ 2007, \apj, 665, 707 

\bibitem[Kiel and Hurley (2009)]{kh09}P. D. Kiel and J. R. Hurley,
  Mon. Not. R. Astron. Soc. 395, 2326 (2009).

\bibitem[King(1962)]{1962AJ.....67..471K} King, I.\ 1962, \aj, 67, 471 

\bibitem[Kopparapu et al.(2008)]{2008ApJ...675.1459K} Kopparapu, R.~K., Hanna, C., Kalogera, V., et al.\ 2008, \apj, 675, 1459-1467 

\bibitem[Kozai(1962)]{1962AJ.....67..591K} Kozai, Y.\ 1962, \aj, 67, 591 


\bibitem[Lee(1995)]{1995MNRAS.272..605L} Lee, H.~M.\ 1995, \mnras, 272, 605 

\bibitem[Lee(1993)]{1993ApJ...418..147L} Lee, M.~H.\ 1993, \apj, 418, 147 

\bibitem[Leigh et al.(2013)]{2013MNRAS.429.2997L} Leigh, N.~W.~C., B{\"o}ker, T., Maccarone, T.~J., \& Perets, H.~B.\ 2013, \mnras, 429, 2997 

\bibitem[Leigh et al.(2014)]{2014MNRAS.441..919L} Leigh, N.~W.~C., Mastrobuono-Battisti, A., Perets, H.~B., {\ Bouml}ker, T.\ 2014, \mnras, 441, 919 

\bibitem[Lidov(1962)]{1962P&SS....9..719L} Lidov, M.~L.\ 1962, \planss, 9, 719 

\bibitem[Lousto \& Zlochower(2008)]{2008PhRvD..77d4028L} Lousto, C.~O., \& Zlochower, Y.\ 2008, \prd, 77, 044028 

\bibitem[{{Lousto} {et~al.}(2012){Lousto}, {Zlochower}, {Dotti}, \&
  {Volonteri}}]{lousto+12}
{Lousto} C.~O., {Zlochower} Y., {Dotti} M., {Volonteri} M., 2012, \prd, 85,
  084015

 \bibitem[Mandel(2016)]{2016MNRAS.456..578M} Mandel, I.\ 2016, \mnras, 456, 578 .

\bibitem[Merritt et al.(2001)]{2001Sci...293.1116M} Merritt, D., Ferrarese, L., \& Joseph, C.~L.\ 2001, Science, 293, 1116 

\bibitem[Merritt(2009)]{2009ApJ...694..959M} Merritt, D.\ 2009, \apj, 694, 959 

\bibitem[Merritt(2010)]{2010ApJ...718..739M} Merritt, D.\ 2010, \apj, 718, 739 

\bibitem[Merritt~(2013)]{david-book}Merritt, D., Dynamics and Evolution of Galactic Nuclei, 2013, Princeton University Press

\bibitem[Miller(2002)]{2002ApJ...581..438M} Miller, M.~C.\ 2002, \apj, 581, 438 

\bibitem[Miller \& Hamilton(2002)]{2002MNRAS.330..232C} Miller, M.~C., \& Hamilton, D.~P.\ 2002, \mnras, 330, 232 


\bibitem[Miller \& Lauburg(2009)]{2009ApJ...692..917M} Miller, M.~C., \& Lauburg, V.~M.\ 2009, \apj, 692, 917 

\bibitem[Miller \& Miller(2015)]{2015PhR...548....1M} Miller, M.~C., \& Miller, J.~M.\ 2015, \physrep, 548, 1 

\bibitem[Morscher et al.(2013)]{2013ApJ...763L..15M} Morscher, M., Umbreit, S., Farr, W.~M., \& Rasio, F.~A.\ 2013, \apjl, 763, L15 

\bibitem[Morscher et al.(2015)]{2015ApJ...800....9M} Morscher, M., Pattabiraman, B., Rodriguez, C., Rasio, F.~A., \& Umbreit, S.\ 2015, \apj, 800, 9 

\bibitem[Neumayer \& Walcher(2012)]{2012AdAst2012E..15N} Neumayer, N., \& Walcher, C.~J.\ 2012, Advances in Astronomy, 2012, 709038 

\bibitem[O'Leary et al.(2009)]{2009MNRAS.395.2127O} O'Leary, R.~M., Kocsis, B., \& Loeb, A.\ 2009, \mnras, 395, 2127 

\bibitem[O'Leary et al.(2016)]{2016arXiv160202809O} O'Leary, R.~M., Meiron, Y., \& Kocsis, B.\ 2016, arXiv:1602.02809 

\bibitem[Pejcha \& Thompson(2015)]{2015ApJ...801...90P} Pejcha, O., \& Thompson, T.~A.\ 2015, \apj, 801, 90 

\bibitem[Perets(2009)]{2009ApJ...698.1330P} Perets, H.~B.\ 2009, \apj, 698, 1330 

\bibitem[Peters(1964)]{1964PhRv..136.1224P} Peters, P.~C.\ 1964, Physical Review, 136, 1224 

 \bibitem[Pfuhl et al.(2011)]{2011ApJ...741..108P} Pfuhl, O., Fritz, T.~K., Zilka, M., et al.\ 2011, \apj, 741, 108 
 
\bibitem[Pfuhl et al.(2014)]{2014ApJ...782..101P} Pfuhl, O., Alexander, T., Gillessen, S., et al.\ 2014, \apj, 782, 101 

\bibitem[Planck Collaboration et al.(2015)]{2015arXiv150201582P} Planck Collaboration, Adam, R., Ade, P.~A.~R., et al.\ 2015, arXiv:1502.01582 

\bibitem[Portegies Zwart \& McMillan(2000)]{2000ApJ...528L..17P} Portegies Zwart, S.~F., \& McMillan, S.~L.~W.\ 2000, \apjl, 528, L17 

\bibitem[Quinlan \& Shapiro(1990)]{1990ApJ...356..483Q} Quinlan, G.~D., \& Shapiro, S.~L.\ 1990, \apj, 356, 483 

\bibitem[Quinlan(1996)]{1996NewA....1...35Q} Quinlan, G.~D.\ 1996, new  astronomy, 1, 35 

\bibitem[Repetto et al.(2012)]{2012MNRAS.425.2799R} Repetto, S., Davies, M.~B., \& Sigurdsson, S.\ 2012, \mnras, 425, 2799 

\bibitem[Repetto \& Nelemans(2015)]{2015MNRAS.453.3341R} Repetto, S., \& Nelemans, G.\ 2015, \mnras, 453, 3341 

\bibitem[Rodriguez et al.(2015)]{2015PhRvL.115e1101R} Rodriguez, C.~L., Morscher, M., Pattabiraman, B., et al.\ 2015, Physical Review Letters, 115, 051101 

\bibitem[Rodriguez et al.(2016a)]{2016arXiv160202444R} Rodriguez, C.~L., Chatterjee, S., \& Rasio, F.~A.\ 2016a, \prd, 93, 084029 

\bibitem[Rodriguez et al.(2016b)]{2016arXiv160404254R} Rodriguez, C.~L., Haster, C.-J., Chatterjee, S., Kalogera, V., \& Rasio, F.~A.\ 2016b, \apjl, 824, L8 




\bibitem[Rossa et al.(2006)]{2006AJ....132.1074R} Rossa, J., van der Marel, R.~P., B{\"o}ker, T., et al.\ 2006, \aj, 132, 1074 

\bibitem[Seth et al.(2008)]{2008ApJ...678..116S} Seth, A., Ag{\"u}eros, M., Lee, D., \& Basu-Zych, A.\ 2008, \apj, 678, 116-130 


\bibitem[Sigurdsson \& Hernquist(1993)]{1993Natur.364..423S} Sigurdsson, S., \& Hernquist, L.\ 1993, \nat, 364, 423 

\bibitem[Spera et al.(2015)]{2015MNRAS.451.4086S} Spera, M., Mapelli, M., \& Bressan, A.\ 2015, \mnras, 451, 4086 

\bibitem[Spitzer~(1987)]{Spitzer}Spitzer, L. \ 1987, Dynamical evolution of Globular Clusters (Princeton: Princeton Univ. Press)

\bibitem[Tsang(2013)]{2013ApJ...777..103T} Tsang, D.\ 2013, \apj, 777, 103 

\bibitem[VanLandingham et al.(2016)]{2016arXiv160404948V} VanLandingham, J.~H., Miller, M.~C., Hamilton, D.~P., \& Richardson, D.~C.\ 2016, arXiv:1604.04948 


\bibitem[Walcher et al.(2006)]{2006ApJ...649..692W} Walcher, C.~J., B{\"o}ker, T., Charlot, S., et al.\ 2006, \apj, 649, 692 

\bibitem[Wang et al.(2015)]{2015MNRAS.450.4070W} Wang, L., Spurzem, R., Aarseth, S., et al.\ 2015, \mnras, 450, 4070 


\end{thebibliography}
  \end{document}